\documentclass[12pt]{article}
\usepackage{a4wide}
\usepackage{amsmath}
\usepackage{amsthm}
\usepackage{amsfonts}
\usepackage{amssymb}

\usepackage{cite}

\DeclareSymbolFont{AMSb}{U}{msb}{m}{n}
\DeclareSymbolFontAlphabet{\Bbb}{AMSb}

\newcommand{\N}{\Bbb{N}}
\newcommand{\Z}{\Bbb{Z}}
\newcommand{\R}{\Bbb{R}}
\newcommand{\C}{\Bbb{C}}

\theoremstyle{plain}
\newtheorem{theorem}{Theorem}
\newtheorem{lemma}[theorem]{Lemma}
\newtheorem{proposition}[theorem]{Proposition}

\theoremstyle{plain}

\newtheorem{assumption}{Assumption}

\DeclareMathOperator{\tr}{tr}

\begin{document}

\thispagestyle{empty}

\vspace*{-80pt} {\hspace*{\fill} Preprint-KUL-TF-2000/02}
\vspace{80pt}

\begin{center} 
  {\LARGE Goldstone Boson Normal Coordinates}\\[25pt]

  {\large T.~Michoel\footnote{Research Assistant of the Fund for
      Scientific Research - Flanders (Belgium)
      (F.W.O.)}\footnotetext{Email: {\tt
        tom.michoel@fys.kuleuven.ac.be}}, A.~Verbeure\footnote{Email:
      {\tt andre.verbeure@fys.kuleuven.ac.be}}
    } \\[25pt]
  {Instituut voor Theoretische Fysica} \\
  {Katholieke Universiteit Leuven} \\
  {Celestijnenlaan 200D} \\
  {B-3001 Leuven, Belgium}\\[25pt]
  {January 20, 2000}\\
  {Final form: June 28, 2000}\\[60pt]
\end{center}

\begin{abstract}\noindent
  The phenomenon of spontaneous symmetry breaking is well known. It is
  known to be accompanied with the appearance of the `Goldstone
  boson'. In this paper we construct the canonical coordinates of the
  Goldstone boson, for quantum spin systems with short range as well
  as long range interactions.
  \\[15pt]
\begin{center}
  {\bf Keywords}
\end{center}
Spontaneous symmetry breaking, Goldstone theorem, plasmon frequency,
long wavelength limits, canonical coordinates.
\end{abstract}
\newpage

\section{Introduction}

As is well known, spontaneous symmetry breakdown (SSB) is one of the
basic phenomena accompanying collective phenomena, such as phase
transitions in statistical mechanics, or ground state excitations in
field theory. SSB is a representative tool for the analysis of many
phenomena in modern physics. The study of SSB goes back to the
Goldstone Theorem \cite{goldstone:1961}, which was the subject of much
analysis. This theorem refers usually to the ground state property
that for short range interacting systems, SSB implies the absence of
an energy gap in the excitation spectrum
\cite{kastler:1966,swieca:1967}.

In this paper we concentrate on the non-relativistic Goldstone
Theorem, and we mean by this spontaneous symmetry breaking of a
continuous symmetry group in condensed matter homogeneous many
particles systems, with short range as well as long range
interactions.

There are many different situations to consider.  For short range
interactions, it is typical that SSB yields a dynamics which remains
symmetric in the thermodynamic limit. At temperature $T=0$, one has as
main characteristics the absence of an energy gap. However for
equilibrium states ($T>0$), SSB is better characterized by bad
clustering properties \cite{martin:1982,fannes:1982b}.

For long range interactions, it is typical that SSB breaks also the
symmetry of the dynamics. This situation has been studied extensively
in the literature. In physics the phenomenon is known as the occurence
of oscillations with energy spectrum taking a finite value
$\epsilon(k\to 0)\not= 0$. Different approximation methods, typical
here is the random phase approximation, yield the computation of these
frequencies. For mean field models, such as the BCS-model
\cite{goderis:1991}, the Overhauser model \cite{broidioi:1991}, a spin
density wave model \cite{broidioi:1991b}, the anharmonic crystal model
\cite{verbeure:1992}, and for the jellium model \cite{broidioi:1993},
one is able to give the rigorous mathematical status of these
frequencies as elements of the spectrum of typical fluctuation
operators \cite{goderis:1989b,goderis:1990}. The typical operators
entering in the discussion are the generator of the broken symmetry
and the order parameter operator. In a physical language they are the
charge density and current density operators. It is proved that their
fluctuation operators form a quantum canonical pair, which decouples
from the other degrees of freedom of the system. As fluctuation
operators are collective operators, they describe the collective mode
accompanying the SSB phenomenon. Hence for long range interacting
systems, we realised mathematically rigorously in these models, the
so-called Anderson theorem \cite{anderson:1958,stern:1966} of
`restoration of symmetry', stating that there exists a spectrum of
collective modes $\epsilon(k\to 0)\not= 0$ and that the mode in the
limit $k\to 0$ is the operator which connects the set of degenerate
temperature states, i.e. `rotates' one ergodic state into an other. We
conjecture that our results of
\cite{goderis:1991,broidioi:1991,broidioi:1991b,verbeure:1992,broidioi:1993}
can be proved for general long range two-body interacting systems as a
universal theorem.

However Anderson did formulate his theorem in the context of the
Goldstone theorem for short range interacting systems, i.e. in the
case $\epsilon(k\to 0)=0$ of absence of an energy gap in the ground
state. Of course one knows that there is no one-to-one relation
between long range interactions and the presence of an energy gap for
symmetry breaking systems (see e.g. \cite{verbeure:1992}). The
imperfect Bose gas and the weakly interacting Bose gas are examples of
long range interacting systems showing SSB, but without energy gap. In
\cite{michoel:1999b} we realise for these boson models the above
described programme of construction of the collective modes operators
of condensate density and condensate current, as normal modes
dynamically independent from the other degrees of freedom of the
system. We consider the whole temperature range, the ground state
included.  In particular the ground state situation is interesting,
because it yields a non-trivial quantum mechanical canonical pair of
conjugate operators, giving an explicit representation of the field
variables of the socalled Goldstone boson.

In this paper we are able to present the analogous proof for general
interacting quantum lattice systems, and hence give {\em a model
  independent construction}.  We construct the fluctuation operators
of the generator of a broken symmetry and of the order parameter and
prove that they form a canonical pair. We prove that this pair is
dynamically independent from the other degrees of freedom of the
system.

In the case of long range interactions, we prove that the appearance
of a plasmon fequency is a natural phenomenon corresponding to the
spectrum of the above mentioned canonical pair. Moreover these
fluctuation operators are normal. Our main contribution here is {\em
  the construction of a canonical order parameter}. Usually there are
many order parameter operators. Therefore the identification of the
right one for the purpose is important.

For short range interactions in the ground state, we find again the
phenomenon of squeezing of the fluctuation operator of the generator
of the broken symmetry. In the literature this is sometimes referred
to the statement that in case of SSB, the broken symmetry behaves like
an approximate symmetry. The amount of squeezing is inversely related
to the anormality of the fluctuation operator of the order parameter,
which itself is directly related to the degree of off-diagonal long
range order. Using an appropriate volume scaling, which is determined
by the long wavelength behaviour of the spectrum, we arrive at the
construction of the Goldstone boson normal coordinates. We consider
this result as a formal step forward, beyond the known analysis of the
Goldstone phenomenon. We repeat that our construction is solely
determined by the long wavelength behaviour of the microscopic energy
spectrum of the system.

Finally, we want to throw the attention of the reader to the direct
open questions which should keep our attention. There is first of all
the problem of SSB of more dimensional symmetries. One should expect a
more dimensional Goldstone boson. There is also the problem of SSB of
non-commutative symmetry groups. An insight in this situation would
certainly contribute to information on the situation of SSB in gauge
theories in relativistic field theory.

\section{Canonical coordinates}\label{can-coord}


\subsection{Introduction}\label{can-coord/intro}
In \cite{goderis:1989b,goderis:1990} a dynamical system of macroscopic
quantum fluctuations is constructed for sufficiently clustering
states. We repeat the main results in order to fix the notation and
refer to the original papers for more details and proofs. The main
issue of this section is the construction of creation and annihilation
operators for this system of macroscopic fluctuation observables. We
start by formulating the systems and the technical settings.

With each $x\in\Z^\nu$ we associate the algebra $\mathcal{A}_x$, a
copy of the matrix algebra $M_N$ of $N\times N$ matrices. For each
$\Lambda\subset\Z^\nu$, consider the tensor product
$\mathcal{A}_\Lambda=\bigotimes_{x\in\Lambda}\mathcal{A}_x$. The
algebra of all local observables is
\begin{equation*}
\mathcal{A}_L = \bigcup_{\Lambda\subset\Z^\nu}\mathcal{A}_\Lambda.
\end{equation*}
The norm closure $\mathcal{A}$ of $\mathcal{A}_L$ is again a
$C^*$-algebra
\begin{equation*}
\mathcal{A} = \overline{\mathcal{A}_L} = \overline{\bigcup_{\Lambda\subset\Z^\nu}
\mathcal{A}_\Lambda},
\end{equation*}
and is considered the algebra of quasi-local observables of our
system.

The group $\Z^\nu$ of space translations of the lattice acts as a
group of *-automorphisms on $\mathcal{A}$ by:
\begin{equation*}
\tau_x: A\in\mathcal{A}_\Lambda \rightarrow \tau_x(A) \in \mathcal{A}_{\Lambda+x}\;,\;\;
x\in\Z^\nu.
\end{equation*}
The dynamics of our system is determined in the usual way by the local
Hamiltonians
\begin{equation*}
H_\Lambda = \sum_{X\subseteq \Lambda}\Phi(X),\;\;\Lambda\subset\Z^\nu
\end{equation*}
with self adjoint $\Phi(X)\in\mathcal{A}_X$ for all $X\subset\Z^\nu$.
The interaction $\Phi$ is supposed to be translation invariant:
\begin{equation*}
\tau_x \Phi(X) = \Phi(X+x).
\end{equation*}
For each $\Lambda\subset\Z^\nu$, the local dynamics $\alpha_t^\Lambda$
is given by
\begin{align*}
  \alpha_t^\Lambda&:\mathcal{A}_\Lambda \rightarrow \mathcal{A}_\Lambda\\
  \alpha_t^\Lambda(A) &=
  e^{itH_\Lambda}Ae^{-itH_\Lambda},\;\;A\in\mathcal{A}_\Lambda.
\end{align*}
If there exists $\lambda>0$ such that
\begin{equation}\label{sh.rg.}
\|\Phi\|_\lambda \equiv \sum_{0\in X} |X|N^{2|X|}e^{\lambda d(X)}\|\Phi(X)\| < \infty,
\end{equation}
with $d(X)=\sup_{x,y\in X}|x-y|$ the diameter of the set $X$ and $|X|$
the number of elements in $X$, then the global dynamics $\alpha_t$ is
well defined as the norm limit of the local dynamics
$\alpha_t^\Lambda$ \cite{bratteli:1996}.

The state $\omega$ is an $(\alpha_t,\beta)$-KMS state which is
supposed to have good spatial clustering expressed by
\begin{equation}\label{cluster}
\sum_{x\in\Z^\nu}\alpha_\omega(|x|)<\infty,
\end{equation}
with $\alpha_\omega$ the following clustering function:
\begin{equation}\label{cluster-f}
\alpha_\omega(d)=\sup_{\Lambda,\Lambda'}\sup_{A\in\mathcal{A}_\Lambda,B\in
\mathcal{A}_{\Lambda'}}\left\{ \frac{1}{\|A\|\|B\|}\Bigl.\left|
\omega(AB)-\omega(A)\omega(B)\right| \;\Bigr|\; d\leq d( \Lambda,\Lambda' )\right\}.
\end{equation}

Through the GNS construction, $\omega$ defines the Gelfand triple
$(\mathcal{H},\pi, \Omega)$, where $\mathcal{H}$ is a Hilbert space,
$\pi$ a *-representation of $\mathcal{A}$ as bounded operators on
$\mathcal{H}$ and $\Omega$ a cyclic vector of $\mathcal{H}$ such that
\begin{equation*}
\omega(A) = (\Omega, \pi(A) \Omega).
\end{equation*}


\subsection{Normal fluctuations}\label{can-coord/norm.fluct.}
Denote by $\Lambda_n$ the cube centered around the origin with edges of length $2n+1$. For
any $A\in\mathcal{A}$, the local fluctuation $F_n(A)$ of $A$ in the state $\omega$ 
is given by
\begin{equation*}
F_n(A)=\frac{1}{|\Lambda_n|^{1/2}}\sum_{x\in\Lambda_n}\bigl(\tau_x A - \omega(A)\bigr).
\end{equation*}

In \cite{goderis:1990} it is proved that under the condition (\ref{cluster}), the central limits exist: for all $A,B\in\mathcal{A}_{L,sa}$ (self-adjoint elements of 
$\mathcal{A}_L$)
\begin{align*}
\lim_{n\to\infty}\omega\bigl( e^{iF_n(A)}e^{iF_n(B)} \bigr) &=
\lim_{n\to\infty}\omega\bigl( e^{iF_n(A+B)}\bigr)e^{-\frac{1}{2}\omega([F_n(A),
F_n(B)])}\\
&= \exp\Bigl\{ -\frac{1}{2}s_\omega\left( A+B,A+B \right)-\frac{i}{2}\sigma_\omega(A,B)
\Bigr\},
\end{align*}
where
\begin{align*}
s_\omega(A,B) &= \lim_{n\to\infty}\;\mathrm{Re}\;\omega\bigl(F_n(A)^*F_n(B)
\bigr) =  \mathrm{Re}\; 
\sum_{x\in\Z^\nu}\bigl( \omega(A^*\tau_xB) - \omega(A^*)\omega(B) \bigr)\\
\sigma_\omega(A,B) &= \lim_{n\to\infty}2\;\mathrm{Im}\;\omega\bigl(F_n(A)^*F_n(B)
\bigr)=-i\sum_x \omega\bigl([A,\tau_x B]).
\end{align*}

Now we are able to introduce the algebra of normal fluctuations of the system
$(\mathcal{A},\mathcal{A}_L,\omega)$. Consider the symplectic space $(\mathcal{A}_{L,sa},
\sigma_{\omega})$. Denote by $W(\mathcal{A}_{L,sa},\sigma_{\omega})$ the CCR-algebra
generated by the Weyl operators $\{W(A)|A\in\mathcal{A}_{L,sa}\}$, satisfying the 
product rule
\begin{equation*}
W(A)W(B)=W(A+B)e^{-\frac{i}{2}\sigma_\omega(A,B)}.
\end{equation*}
The central limit theorem fixes a representation of this CCR-algebra in the following way.
For each $A\in\mathcal{A}_{L,sa}$ the limits
$\lim_{n\to\infty}\omega\bigl(e^{iF_n(A)}\bigr)$
define a quasi-free state $\tilde\omega$ 
of  the CCR-algebra $W(\mathcal{A}_{L,sa},\sigma_{\omega})$ by
\begin{equation*}
\tilde\omega\bigl( W(A) \bigr)=e^{-\frac{1}{2}s_\omega(A,A)}.
\end{equation*}
Moreover if $\gamma$ is a *-automorphism of $\mathcal{A}$ leaving $\mathcal{A}_L$
invariant, commuting with the space translations and leaving the state $\omega$ invariant,
then $\tilde\gamma$ given by
\begin{equation}\label{gamma}
\tilde\gamma(W(A))=W(\gamma(A))
\end{equation}
defines a quasi-free *-automorphism of $W(\mathcal{A}_{L,sa},\sigma_{\omega})$.

The quasi-free state $\tilde\omega$ induces a GNS-triplet 
$(\tilde\mathcal{H},\tilde\pi,\tilde\Omega)$ and
yields a von Neumann algebra
\begin{equation*}
\tilde\mathcal{M} = \tilde\pi\bigl( W(\mathcal{A}_{L,sa},\sigma_{\omega}) \bigr)''.
\end{equation*}
This algebra will be called the {\em algebra of normal (macroscopic) fluctuations}.

By the fact that the representation $\tilde\pi$ is regular, we can define boson fields
$F_0(A)$ given by $\tilde\pi(W(A))=e^{i F_0(A)}$, and satisfying
\begin{equation*}
[F_0(A),F_0(B)]=i\sigma_\omega(A,B).
\end{equation*}
Through the relation
\begin{equation*}
\lim_{n\to\infty}\omega\bigl(e^{iF_n(A)}\bigr) = \tilde\omega\bigl(e^{iF_0(A)}\bigr),
\end{equation*}
we are able to identify  the macroscopic fluctuations of the system  
$(\mathcal{A},\omega)$ with the boson field $F_0(\cdot)$:
\begin{equation*}
\lim_{n\to\infty}F_n(A)=F_0(A).
\end{equation*}

Let $(\mathcal{H},\pi,\Omega)$ be the GNS-triplet induced by the state $\omega$ and
consider the sesquilinear form $\langle\cdot,\cdot\rangle_0$ on $\mathcal{H}$ with domain
$\pi(\mathcal{A}_L)\Omega$ which we simply denote by $\mathcal{A}_L$:
\begin{align}
\langle A,B\rangle_0 &= s_\omega(A,B)+\frac{i}{2}\sigma_\omega(A,B)\label{sesq-form}
=\sum_{x\in\Z^\nu} \bigl( \omega(A^*\tau_xB) - \omega(A^*)\omega(B) \bigr)\nonumber.
\end{align}
We call $A$ and $B$ in $\mathcal{A}_L$ equivalent, denoted $A\equiv_0 B$ if
$\langle A-B,A-B\rangle_0 = 0$.
The following important result holds:
\begin{equation}\label{equiv}
A\equiv_0 B
\Leftrightarrow
\tilde\pi\left( W(A) \right) = \tilde\pi\left( W(B) \right).
\end{equation}
This is the property of \emph{coarse graining}: different micro observables yield the same
macroscopic fluctuation operator.

Denote by $[\mathcal{A}_L]$ the equivalence classes of $\mathcal{A}_L$ for the equivalence
relation $\equiv_0$. The form $\langle\cdot,\cdot\rangle_0$ is a scalar product on 
$[\mathcal{A}_L]$.
Denote by $\mathcal{K}_\omega$ the Hilbert space obtained as the completion of 
$[\mathcal{A}_L]$. Clearly $s_\omega$ and $\sigma_\omega$ extend continuously to 
$\mathcal{K}_\omega$. Denote by $\mathcal{K}_\omega^{Re}$ the real subspace of 
$\mathcal{K}_\omega$ generated by $[\mathcal{A}_{L,sa}]$. Now one considers the
CCR-algebra $W\left(\mathcal{K}_\omega^{Re},\sigma_\omega\right)$ in the 
same representation induced by the state $\tilde\omega$, and 
 one has the following equality:
\begin{equation*}
\tilde\mathcal{M} = \tilde\pi\bigl(
W(\mathcal{K}_\omega^{Re},\sigma_\omega)\bigr)''.
\end{equation*}


\subsection{Reversible dynamics of fluctuations}
Property (\ref{gamma}) is not directly applicable with $\gamma=\alpha_t$, because with this
choice it is not clear, and generally not true that $\alpha_t\mathcal{A}_L\subseteq 
\mathcal{A}_L$. Nevertheless, since $\alpha_t F_n(A)=F_n(\alpha_t A)$ one is tempted to
define the dynamics $\tilde \alpha_t$ of the fluctuations by the formula
\begin{equation*}
\tilde \alpha_t F_0(A) = F_0(\alpha_t A).
\end{equation*}
The non-trivial point in this formula is that it is unclear whether the central limit of the
non-local observable $\alpha_t A$ exists or not. Furthermore if $F_0(\alpha_t A)$
exists it remains to prove that $(\tilde \alpha_t)_t$ defines a weakly continuous group of
*-automorphisms on the fluctuation algebra $\tilde\mathcal{M}$. 

In \cite{goderis:1990} it is shown that if the interaction $\Phi$ is of short range, i.e. 
if $\Phi$ satisfies condition (\ref{sh.rg.}), then
for all  $A\in[\mathcal{A}_L]$, one has that for all $t\in\R$,
$\alpha_tA\in\mathcal{K}_\omega$ and if
$A\in[\mathcal{A}_{L,sa}]$ then $\alpha_tA\in\mathcal{K}_\omega^{Re}$. $W(\alpha_tA)$ is a
well defined element of $\tilde\mathcal{M}$ and as
\begin{equation*}
W(\alpha_tA) = e^{iF_0(\alpha_t A)},\;A\in[\mathcal{A}_{L,sa}]
\end{equation*}
the fluctuation $F_0(\alpha_t A)$ exists for all $t\in\R$.

The map $U_t:[\mathcal{A}_L]\rightarrow\mathcal{K}_\omega$, $U_tA=\alpha_tA$ is a well
defined linear operator on the Hilbert space
$\left(\mathcal{K}_\omega,\langle\cdot,\cdot\rangle_0\right)$ 
extending to a unitary operator for all $t\in\R$. The map $t\rightarrow U_t$ is a
strongly continuous one-parameter group, and for all elements $A\in\mathcal{K}_\omega^{Re}$
we can define $\tilde \alpha_t W(A) = W(U_t A)$.
Then $\tilde \alpha_t$ extends to a weakly continuous one-parameter group of
*-automorphisms of $\tilde\mathcal{M}$.

Moreover it is shown that if the microsystem is in an equilibrium state, then also
the macro system of fluctuations is in an equilibrium state for the dynamics constructed
in the previous theorem, i.e. the notion of equilibrium is preserved under the operation of
coarse graining induced by the central limit. In particular,
if $\omega$ is an $\alpha_t$-KMS state of $\mathcal{A}$ at $\beta>0$, then $\tilde\omega$ is
an $\tilde\alpha_t$-KMS state of the von Neumann algebra $\tilde\mathcal{M}$ at the same
temperature.


\subsection{Canonical coordinates}\label{can-coord/can.coord.}
Now we proceed to the explicit construction of creation and annihilation operators 
of fluctuations in the algebra $\tilde\mathcal{M}$. For product states this construction
can be found in \cite{goderis:1989}. 
Here we work out the construction for the most general system. 

From the definition of $\mathcal{K}_\omega^{Re}$ and $\mathcal{K}_\omega$ we can write
\begin{equation*}
\mathcal{K}_\omega=\mathcal{K}_\omega^{Re} + i\mathcal{K}_\omega^{Re}.
\end{equation*}
Let * be the operation on $\mathcal{K}_\omega$ defined by
\begin{equation*}
A^* = (A_1+iA_2)^*=A_1-iA_2,\;A_1,A_2\in\mathcal{K}_\omega^{Re}.
\end{equation*}
Clearly for $X\in\mathcal{A}_L$ one has $[X]^*=[X^*]$ and it follows from the properties
of $U_t$ (see above) that
\begin{equation*}
(U_tA)^* = U_t A^*
\end{equation*}
for all $A\in\mathcal{K}_\omega$.

Let $\mathcal{D}$ denote the set of infinitely differentiable functions on $\R$ with compact
support. $\mathcal{D}$ is dense in $\mathcal{C}_0(\R)$, the continuous functions vanishing
at $\infty$, for the supremum norm. If $\hat f\in\mathcal{D}$ then the inverse Fourier
transform
\begin{equation*}
f(z)=\int_{-\infty}^{+\infty}d\lambda\hat f(\lambda)e^{i\lambda z}
\end{equation*}
is an entire analytic function. If supp $\hat f \in [-R,R]$ then it follows from the
theorem of Paley-Wiener \cite{bratteli:1996} that for all $n\in\N$ there 
exists a constant $C_n$ such that
\begin{equation*}
|f(z)|\leq C_n (1+|z|)^{-n}e^{R|\mathrm{Im}z|}.
\end{equation*}

Let $U_t=e^{it\tilde h}=\int e^{it\lambda}d\tilde E_\lambda$ be the 
spectral resolution of the
unitary group $U_t$ and for $A\in\mathcal{K}_\omega$, $f\in L^1(\R)$ denote
\begin{align*}
A(f) &= \int_{-\infty}^{+\infty} dt f(t)U_tA
=\int_{-\infty}^{+\infty}\hat f(-\lambda)d\tilde E_\lambda A
= \hat f(-\tilde h)A.
\end{align*}
Clearly one has $A(f)^*=A^*(\bar f)$.

Let $W$ be an open set in $\R$ and let $\tilde E_W=\int_W d\tilde E_\lambda$ be the spectral projection
onto the spectral subspace $\mathcal{K}_W$. It follows from the spectral theory
\cite{bratteli:1996,arveson:1974} that $\mathcal{K}_W$ is generated by the set
\begin{equation*}
\{ A(f) | A\in\mathcal{K}_\omega,\; f\in \mathcal{D},\; \mathrm{supp}\;\hat f \subset W\}.
\end{equation*}

Finally for $A\in\mathcal{K}_\omega$ denote the associated spectral measure by
\begin{equation*}
d\tilde\mu_A(\lambda)=\langle A,d\tilde E_\lambda A\rangle_0
\end{equation*}
and its spectral support $\Delta_A$
\begin{equation}\label{eq:spectr-supp}
\Delta_A=\{\lambda\in\R \;|\; \tilde\mu_A([\lambda-\epsilon,\lambda+\epsilon])>0 \;\forall
\epsilon>0\}.
\end{equation}
It is easy to see that $\Delta_A$ is also given by
$\Delta_A=\{\lambda\in\R \;|\; \hat f(\lambda)=0,\;\forall\hat
f\in\mathcal{D}\;\mathrm{such\;that}\;A(f)=0 \}$.
From this expression and $\Bar{\Hat{f}}(\lambda)=\Hat{\Bar{f}}(-\lambda)$ it follows that
$\Delta_{A^*}=-\Delta_A$,
and from the same argument one also has
\begin{equation}\label{E-adjoint}
\tilde E_+ A^* = (\tilde E_- A)^*
\end{equation}
where $\tilde E_+ = \tilde E_{(0,+\infty)}$ and $\tilde E_-=\tilde E_{(-\infty,0)}$ are
the projections onto positive, respectively negative energy.

\begin{lemma}\label{le:kms}
Let $\omega$ be an ($\alpha_t,\beta$)-KMS state on the algebra $\mathcal{A}$. For all 
$A\in\mathcal{K}_\omega$, $\hat f\in\mathcal{D}$
\begin{equation*}
\int\hat f(\lambda)d\mu_A(\lambda) = \int\hat f(\lambda)e^{\beta\lambda}d\mu_{A^*}
(-\lambda).
\end{equation*}
\end{lemma}
\begin{proof}
Follows from the KMS-properties of $\tilde\omega$.
\end{proof}

Let $\mathcal{K}_{\omega,0}^{Re}=\tilde E_0\mathcal{K}_\omega^{Re}$ and 
$\mathcal{K}_{\omega,1}^{Re}=(\tilde E_+ +\tilde E_-)\mathcal{K}_\omega^{Re}$. 
Define the operator $J$ on $\mathcal{K}_{\omega,1}^{Re}$ by
\begin{equation}\label{compl-struct}
J=i(\tilde E_+ - \tilde E_-).
\end{equation}
From (\ref{E-adjoint}) one has for all $A\in\mathcal{K}_{\omega,1}^{Re}$,
$(JA)^* = JA^*$
and thus $J\mathcal{K}_{\omega,1}^{Re}\subseteq\mathcal{K}_{\omega,1}^{Re}$.
\begin{proposition}\label{prop-compl-struct}
The operator $J$ defined above is a complex structure on the symplectic space
$(\mathcal{K}_{\omega,1}^{Re},\sigma_\omega)$:
\begin{enumerate}
\item $J^2 = -1$

\item $\sigma_\omega(A,JB) = -\sigma_\omega(JA,B),\;\;\;A,B\in\mathcal{K}_{\omega,1}^{Re}$

\item $\sigma_\omega(A,JA)>0,\;\;\;0\not=A\in\mathcal{K}_{\omega,1}^{Re}$
\end{enumerate}
\end{proposition}
\begin{proof}
From the definition of $J$ and $\sigma_\omega=2\;\mathrm{Im}\;\langle\cdot,\cdot
\rangle_0$, $(i)$ and $(ii)$ are trivially satisfied. Now we prove $(iii)$.
Let $\mathcal{E}$ be the set of real functions $f$ such that $\hat f\in\mathcal{D}$ and 
$0\notin$ supp $\hat f$. By the spectral theory, the set generated by
$\{A(f)| A\in\mathcal{K}_{\omega,1}^{Re},\; f\in\mathcal{E} \}$
is dense in $\mathcal{K}_{\omega,1}^{Re}$.

Take such an element $A(f)$. Using the previous lemma one computes
\begin{align*}
\langle  \tilde E_-A(f), \tilde E_-A(f)\rangle_0 &= \int |\hat f(\lambda)|^2 
\chi_{(-\infty,0)}(\lambda)
d\mu_A(\lambda)
=\int |\hat f(-\lambda)|^2 e^{-\beta\lambda} \chi_{(0,\infty)}(\lambda)
d\mu_A(\lambda)\\
&=\langle  \tilde E_+A(f), e^{-\beta \tilde h}\tilde E_+A(f)\rangle_0.
\end{align*}
Because $\tilde E_+,\tilde E_-$ are projections and $e^{-\beta\tilde h}=
\int e^{-\beta\lambda}d\tilde E_\lambda$ is
bounded on $\tilde E_+\mathcal{K}_{\omega,1}^{Re}$, this relation holds for all $B\in\mathcal{K}_{\omega,1}^{Re}$.
Using this property one has 
\begin{align*}
\sigma_\omega(A,JA) &= -2i\;\mathrm{Im}\; \langle A,JA \rangle_0
= 2 \left( \langle \tilde E_+A,\tilde E_+A \rangle_0 - \langle \tilde E_-A,\tilde E_-A \rangle_0 \right)\\
&= 2\int_0^{\infty} (1-e^{-\beta \lambda})\langle A,d\tilde E_\lambda A \rangle_0\geq 0.
\end{align*}
The strict inequality holds because the spectral measure $d\tilde\mu_A(\lambda)$ is 
regular and $\tilde E_0A=0$.
\end{proof}

The existence of a complex structure $J$ yields the existence of creation and annihilation
operators
\begin{equation}\label{eq:cr-ann-ops}
a^\pm_0(A) = \frac{F_0(A)\mp iF_0(JA)}{\sqrt 2}
\end{equation}
for all $A\in\mathcal{K}_{\omega,1}^{Re}$. They satisfy the property
\begin{equation*}
a^\pm_0(JA)=\pm i a^\pm_0(A).
\end{equation*}


\subsection{Normal modes}
\label{can-coord/2.5}

Consider a given microscopic observable $A$ such that $[A]\in
\mathcal{K}_{\omega,1}^{Re}$, i.e. such that $F_0(A)$ evolves
non-trivially under the dynamics $\tilde\alpha_t$. For simplicity we
will denote $A=[A]$. We will construct the normal modes corresponding
to the macroscopic fluctuations of the observable $A$.

In order to make clear the idea we will first make the simplyfying
assumption that the spectral measure $d\tilde\mu_A(\lambda)$ consists
of two $\delta$-peaks, at $\pm \epsilon_A$, with $\epsilon_A >0$.
Afterwards we will show how to extend the construction to more general
(absolutely continuous) measures $d\tilde\mu_A$. Notice also that the
prototype examples of systems with normal fluctuations, i.e. mean
field systems, have a discrete energy spectrum and therefore obey the
$\delta$-peak assumption (see section \ref{long-range} for an explicit
example).

\begin{lemma}\label{le:f-correl}
  For $\hat f\in\mathcal{D}$ and $[A]\in\mathcal{K}_{\omega,1}^{Re}$,
\begin{equation*}
\int \hat f(\lambda) d\tilde\mu_A(\lambda)=\int_0^\infty \bigl( \hat f(\lambda)+
\hat f(-\lambda)e^{-\beta\lambda}\bigr)d\tilde\mu_A(\lambda),
\end{equation*}
and for $\hat f( \tilde h)A\in\mathcal{K}_{\omega}^{Re}$ (i.e. $f(t)$
real),
\begin{equation*}
\tilde \omega\Bigl( F_0\bigl(\hat f(\tilde h)A\bigr)^2 \Bigr)=
\int |\hat f(\lambda)|^2 d\tilde\mu_A(\lambda).
\end{equation*}
\end{lemma}
\begin{proof}
  This is a simple computation and application of Lemma
  \ref{le:kms}.
\end{proof}

It will turn out to be more natural to work in terms of the following
measure: for $\lambda>0$
\begin{equation*}
dc_A(\lambda)\equiv 2\frac{1-e^{-\beta\lambda}}{\lambda}d\tilde\mu_A(\lambda),
\end{equation*}
and $0$ otherwise, such that by Lemma \ref{le:f-correl}
\begin{equation*}
c_A\equiv \int_0^\infty dc_A(\lambda)=\int_{-\infty}^{+\infty}\frac{1-e^{-\beta\lambda}}
{\lambda}d\tilde\mu_A(\lambda)=\beta\bigl(F_0(A),F_0(A)\bigr)_\sim
\end{equation*}
is the well known Duhamel two point function, or canonical
correlation. In the sequel, $c_A$ will act as a \emph{quantization
  parameter} or Planck's constant for the normal modes corresponding
to the fluctuations of $A$.

The assumption on the spectral measure of the fluctuations of $A$ then
amounts to the assumption that there exists $\epsilon_A >0$ such that
\begin{equation}\label{eq:assum-A}
dc_A(\lambda) = c_A \delta(\lambda-\epsilon_A)d\lambda.
\end{equation}

The ``position'' operator $Q_0(A)$ and ``momentum'' operator $P_0(A)$
of the normal mode are now defined by
\begin{align*}
  Q_0(A)\equiv F_0(A) && P_0(A)\equiv F_0(i\tilde h^{-1}A).
\end{align*}
Obviously $P_0(A)$ is well defined because of the assumption
(\ref{eq:assum-A}).

The following proposition justifies the name \emph{normal mode}:
\begin{proposition}\label{pr:norm-mode-A}
  The pair $\bigl(Q_0(A),P_0(A) \bigr)$ forms a quantum canonical
  pair,
\begin{equation*}
\bigl[ Q_0(A),P_0(A) \bigr] = i c_A,
\end{equation*}
satisfying the equations of motion of a free quantum harmonic
oscillator with frequency $\epsilon_A$:
\begin{align*}
  \tilde\alpha_t Q_0(A) &= Q_0(A) \cos\epsilon_A t + \epsilon_A
  P_0(A) \sin\epsilon_A t\\
  \tilde\alpha_t P_0(A) &= -\frac{1}{\epsilon_A}Q_0(A)\sin\epsilon_A t
  + P_0(A) \cos\epsilon_A t.
\end{align*}
The $(\tilde\alpha_t,\beta)$-KMS property of $\tilde\omega$ is
expressed by
\begin{equation*}
\tilde\omega\left( Q_0(A)^2 \right) =\epsilon_A^2 
\tilde\omega\left( P_0(A)^2 \right)
=\frac{c_A\epsilon_A}{2}\coth\frac{\beta\epsilon_A}{2}.
\end{equation*}
\end{proposition}
\begin{proof}
  By the KMS property of $\tilde\omega$
\begin{equation*}
\sigma\bigl( F_0(A),F_0(i\tilde h^{-1}A)\bigr)=\int (1-e^{-\beta\lambda})\lambda^{-1}
d\tilde\mu_A(\lambda)=c_A.
\end{equation*}
Lemma \ref{le:f-correl} and assumption (\ref{eq:assum-A}) yield
\begin{equation*}
\tilde\omega\left( Q_0(A)^2 \right) =\epsilon_A^2 
\tilde\omega\left( P_0(A)^2 \right)
=\frac{c_A\epsilon_A}{2}\coth\frac{\beta\epsilon_A}{2}.
\end{equation*}
A similar computation yields $\langle \epsilon_A JA - i\tilde h A,
\epsilon_A JA - i\tilde h A\rangle_A=0$, and by the equivalence
relation (equation (\ref{equiv})), $F_0(i\tilde h A)=\epsilon_A
F_0(JA)$, and by exponentiation:
\begin{equation*}
\tilde\alpha_t F_0(A)=F_0(e^{\epsilon_A tJ}A);
\end{equation*}
$J^2=-1$ yields
\begin{equation*}
\tilde\alpha_t F_0(A)= F_0(A)\cos \epsilon_A t + F_0(JA)\sin \epsilon_A t.
\end{equation*}
As above one shows that by the equivalence relation (\ref{equiv}),
\begin{equation*}
F_0(i\tilde h^{-1} A)=\epsilon_A^{-1} F_0(JA)
\end{equation*}
yielding the equations of motion as stated in the proposition.
\end{proof} 
The creation and annihilation
operators corresponding to this harmonic mode are simply the creation
and annihilation operators defined in (\ref{eq:cr-ann-ops}), although
it is customary to rescale them with $\sqrt{\epsilon_A}$, i.e.
\begin{equation*}
\frac{1}{\sqrt{\epsilon_A}}a^\pm_0(A)=\frac{Q_0(A)\mp i \epsilon_A P_0(A)}
{\sqrt{2\epsilon_A}}.
\end{equation*}

Let us now consider how this situation can be extended to the more
general case where the measure $d\tilde\mu_A(\lambda)$ has some
spectral support $\Delta_A$ (see (\ref{eq:spectr-supp})). To avoid
problems at energy $\lambda=0$, we assume $\Delta_A$ to be bounded
away from $0$, i.e. there exists $\epsilon_A>0$ such that
\begin{equation*}
\Delta_A^+ \equiv \Delta_A \cap \R^+ \subseteq [\epsilon_A,+\infty).
\end{equation*}
Remark that $\Delta_A^+$ is the support of the measure
$dc_A(\lambda)$. In this case we can safely assume this measure to be
absolutely continuous, i.e.
\begin{equation*}
dc_A(\lambda)=c_A(\lambda)d\lambda.
\end{equation*}
Lemma \ref{le:f-correl} yields
\begin{equation*}
\tilde\omega\bigl(F_0(A)^2\bigr)=\int_{\Delta_A^+}\frac{c_A(\lambda)\lambda}{2}\coth
\frac{\beta\lambda}{2}d\lambda.
\end{equation*}
It is easily seen that instead of a single mode
$\bigl(Q_0(A),P_0(A)\bigr)$ one can construct in this situation a
continuous family of harmonic modes, i.e. two operator valued
distributions
\begin{equation*}
\left\{ \bigl(Q_{0,A}(\lambda),P_{0,A}(\lambda)\bigr)\;|\; \lambda \in \Delta_A^+ \right\},
\end{equation*}
such that
\begin{align*}
  \bigl[ Q_{0,A}(\lambda),P_{0,A}(\lambda')\bigr] &=
  ic_A(\lambda)\delta(\lambda-\lambda')\\
  \tilde\omega\bigl(Q_{0,A}(\lambda)^2\bigr) &=\lambda^2
  \tilde\omega\bigl(P_{0,A}(\lambda)^2\bigr)
  =\frac{c_A(\lambda)\lambda}{2}\coth\frac{\beta\lambda}{2}\\
  \tilde\alpha_t Q_{0,A}(\lambda) &= Q_{0,A}(\lambda) \cos\lambda t +
  \lambda
  P_{0,A}(\lambda)\sin\lambda t\\
  \tilde\alpha_t P_{0,A}(\lambda) &=-\frac{1}{\lambda}
  Q_{0,A}(\lambda) \sin\lambda t + P_{0,A}(\lambda)\cos\lambda t.
\end{align*}
One identifies
\begin{align*}
  F_0(A)=Q_0(A)=\int_{\Delta_A^+}Q_{0,A}(\lambda)d\lambda &&
  F_0(i\tilde h^{-1}A)=P_0(A)
  =\int_{\Delta_A^+}P_{0,A}(\lambda)d\lambda.
\end{align*}
Remark that due to the spectral gap $P_0(A)$ is well defined and that
by the spectral theory \cite{arveson:1974}, $Q_{0,A}(\lambda)$ can be
arbitrarily well approximated \cite[Proposition 3.2.40
]{bratteli:1996} by a sequence of operators $F_0(A(f_i))$, where $\hat
f_i \in\mathcal{D}$ is a sequence converging to a double $\delta$-peak
in $\pm\lambda$.

The content of this paper is to apply the construction of Proposition
\ref{pr:norm-mode-A} to the situation of spontaneous breaking of a
continuous symmetry, where we take for $A$ the symmetry generator
(i.e.  the ``charge'' operator). The normal modes corresponding to the
fluctuations of the symmetry generator as constructed above then yield
a rigorous mathematical representation of the collective modes
accompanying the spontaneous symmetry breaking (SSB), i.e. of the
\emph{Goldstone bosons}.

There are two distinct situations to consider, either the system with
SSB has a gap in the energy spectrum, or it has not. The former
situation is typically connected with long range interactions, the
latter with short range interactions. Both situations introduce
specific problems that make Proposition \ref{pr:norm-mode-A} not
directly applicable as such.

Long range interacting systems in general do not possess a
well-defined time evolution in the thermodynamic limit. Therefore one
is restricted to studying specific models. In section \ref{long-range}
we study a prototype model of a long range interacting system with a
well-defined time evolution and a spectral gap, i.e. a mean field
system. These systems have normal fluctuations, hence one can apply
Proposition \ref{pr:norm-mode-A} directly.

The presence of SSB in short range interacting systems is
characterized by either bad clustering properties (for temperature
$T>0$) or the absence of a spectral gap ($T=0$).  This is the content
of the Goldstone Theorem (see section \ref{short-range} and references
\cite{landau:1981, wreszinski:1987} for more details). Therefore these
systems do not have normal fluctuations as defined in this section,
i.e. there is \emph{off diagonal long range order} in the system.  For
the systems we are interested in, this is a statement that applies to
momentum $k=0$ only, and one goes around this problem by working with
the $k$-mode fluctuations, $k\not= 0$,
\begin{equation*}
F_{n,k}(A)=\frac{1}{|\Lambda_n|^{1/2}}\sum_{x\in\Lambda_n}\bigl( \tau_x A - \omega(A)\bigr)
\cos k.x.
\end{equation*}
These fluctuation operators will be shown to be normal and it will
also be shown that in the ground state ($T=0$) one can recover the
situation of Proposition \ref{pr:norm-mode-A} in a properly scaled
limit $k\to 0$. This is the content of section \ref{short-range}.

\section{Long range interactions}\label{long-range}


\subsection{Introduction}
In this section we study symmetry breaking systems whose Hamiltonian
has a gap in the ground state. These systems typically have long range
interactions, but since there is no general criterium whether a long
range interacting system has a spectral gap or not, and since an
infinite volume time evolution in general may not exist for these
systems (see condition (\ref{sh.rg.})), we restrict ourself to mean
field systems which are long range interacting systems with a well
defined time evolution in the thermodynamic limit and with a spectral
gap.  For the sake of clarity we consider an explicit example, namely
the strong coupling BCS-model for superconductivity.  Similar results
as the ones presented here have already been obtained for different
other mean field models \cite{broidioi:1991,broidioi:1991b}, and for
the jellium model \cite{broidioi:1993}, albeit by different methods.
Moreover our main contribution in this section is {\em the
  construction of a canonical order parameter}.

The Hamiltonian for the strong coupling BCS-model is given by 
\cite{thirring:1967, thirring:1968}
\begin{equation*}
H_N=\epsilon\sum_{i=-N}^N \sigma_i^z - \frac{1}{2N+1}\sum_{i,j=-N}^N \sigma_i^+
\sigma_j^-\;,\;\epsilon<\frac{1}{2}
\end{equation*}
where $\sigma^z,\sigma^\pm$ are the usual ($2 \times 2$) Pauli matrices.
$H_N$ acts on the Hilbert space $\otimes_{i=-N}^N\C_i^2$.

The solutions of the KMS equation are given by the product states 
$\omega_\lambda=\omega_{\rho_\lambda}$
on the infinite tensor product algebra $\mathcal{A}=\otimes_{i=-\infty}^\infty (M_2)_i$ of
the system; $\rho_\lambda$ is a ($2\times 2$) density matrix, given by the solutions of the
gap equation
\begin{equation*}
\rho_\lambda = \frac{e^{-\beta h_\lambda}}{\tr e^{-\beta h_\lambda}}\;,\;\;
\lambda=\tr\rho_\lambda \sigma^- = \omega_\lambda(\sigma^-)\;,\;\;
h_\lambda=\epsilon\sigma^z-\lambda\sigma^+ -\bar\lambda\sigma^-.
\end{equation*}
This is easily turned into the equation for $\lambda$:
\begin{equation}\label{BCSgap}
\lambda\left( 1-\frac{\tanh \beta\mu}{2\mu} \right)=0
\end{equation}
with $\mu=(\epsilon^2+|\lambda|^2)^{1/2}$. Clearly, this equation has always the solution
$\lambda=0$, describing the so-called normal phase. We are interested in the 
solutions $\lambda\not=0$ which exist in the case $\beta >\beta_c$ where
$\beta_c$ is determined by the equation
$\tanh\beta_c\epsilon = 2\epsilon$.
These solutions $\lambda\not= 0$ are understood to describe the superconducting phase.
Remark that if $\lambda\not=0$ is a solution of (\ref{BCSgap}), then for all
$\phi\in[0,2\pi)$, $\lambda e^{i\phi}$ is a solution as well. There is an infinite
degeneracy of the states for the superconducting phase. The degeneracy is due to the
breaking of the gauge symmetry. As
$\sigma^z = \sigma^+\sigma^- - \sigma^-\sigma^+$
it is clear that the Hamiltonian $H_N$ is invariant under the continuous gauge 
transformations automorphism group $\mathcal{G}=\{\gamma_\phi | \phi\in [0,2\pi)\}$ of 
$\mathcal{A}$
\begin{equation*}
\gamma_\phi: \sigma^+_i \rightarrow \gamma_\phi(\sigma^+_i)=e^{-i\phi}\sigma^+_i.
\end{equation*}
However the solutions $\omega_\lambda$ are not invariant for this symmetry transformation,
because:
\begin{equation}\label{sym-br}
\omega_\lambda (\gamma_\phi(\sigma^+_i))=e^{-i\phi}\omega_\lambda(\sigma^+_i)\not=
\omega_\lambda(\sigma^+_i).
\end{equation}
The gauge symmetry of the system is spontaneously broken. Remark  that $h_\lambda$ is
no longer invariant under the symmetry transformation, this is a typical feature of long
range interacting systems.
From (\ref{sym-br}) it follows also that
$\omega_\lambda \circ \gamma_\phi = \omega_{\lambda e^{i\phi}}$,
i.e. one solution $\omega_\lambda$ is transformed into another solution 
$\omega_{\lambda e^{i\phi}}$ by the gauge transformation $\gamma_\phi$.

The gauge group $\mathcal{G}$ is not implemented by
unitaries in any of the representations induced by the solutions $\omega_\lambda$. 
Locally however, the gauge transformation $\gamma_\phi$ is implemented by
unitaries: take any finite set $\Lambda$ of indices, then
\begin{equation*}
\gamma_\phi \Bigl( \prod_{i\in\Lambda}\sigma_i^- \Bigr)= \bigl(U^\Lambda_\phi\bigr)^*
\Bigl( \prod_{i\in\Lambda}\sigma_i^- \Bigr) U^\Lambda_\phi
\end{equation*}
where
\begin{equation*}
U^\Lambda_\phi = e^{\frac{i}{2}\phi Q_\Lambda}\;,\;\;Q_\Lambda=\sum_j  \sigma^z_j.
\end{equation*}
The operator $Q_\Lambda$ is called the local charge or symmetry generator and $\sigma^z$ 
the charge density or symmetry generator density.


\subsection{Canonical coordinates of the Goldstone mode}
Next we introduce the algebra of fluctuations and show how the
Goldstone mode operators are to be defined in a canonical way. The
relation between symmetry breaking and quantum fluctuations in the
strong coupling BCS model has been studied before in
\cite{goderis:1991}.  This analysis is here extended.

Per lattice site $j\in\Z$ one has the local algeba of observables, the
real ($2\times 2$) matrices, $M_2$, generated by the Pauli matrices.
As state we consider a particular equilibrium state $\omega_\lambda$
with $\beta>\beta_c$ which reduces per lattice point to the trace
state $\omega_\lambda(A)=\tr \rho_\lambda A$, $A\in M_2$.  Because of
the product character of the algebra, the state and the time
evolution, it is sufficient to consider fluctuations of one-point
observables. Locally the fluctuation of $A$ in the state
$\omega_\lambda$ is:
\begin{equation*}
F_N(A)=\frac{1}{(2N+1)^{1/2}}\sum_{i=-N}^N \bigl(A_i - \rho_\lambda(A)\bigr)\;,\;\;
A\in M_2.
\end{equation*}
The commutator of two fluctuations is a mean, indeed:
\begin{equation*}
\bigl[F_N(A),F_N(B)\bigr]=\frac{1}{2N+1}\sum_{i=-N}^N \bigl([A,B]\bigr)_i.
\end{equation*}
For $A,B\in M_2$ define
\begin{align*}
  s_\lambda(A,B) &= \mathrm{Re}\; \rho_\lambda\Bigl( \bigl(A -
  \rho_\lambda(A)\bigr)
  \bigl(B - \rho_\lambda(B)\bigr) \Bigr)\\
  \sigma_\lambda(A,B) &= \mathrm{Im}\; \rho_\lambda\Bigl( \bigl[A -
  \rho_\lambda(A)\bigr] \bigl[B - \rho_\lambda(B)\bigr] \Bigr) =
  -i\rho_\lambda\bigl([A,B]\bigr).
\end{align*}
Clearly $(M_{2,sa},\sigma_\lambda)$ is a symplectic space and
$s_\lambda$ is a symmetric positive bilinear form on $M_{2,sa}$.

Because $\rho_\lambda$ is time invariant, $\rho_\lambda
\circ\alpha_t=\rho_\lambda$ and because the evolution $\alpha_t$ is
local, $\alpha_t:M_{2,sa} \rightarrow M_{2,sa}$, one has that
$\alpha_t$ is a symplectic operator on $(M_{2,sa},\sigma_\lambda)$:
for all $t\in\R$
\begin{equation*}
\sigma_\lambda(\alpha_t A,\alpha_t B)=\sigma_\lambda(A,B).
\end{equation*}
The structure $(M_{2,sa},\sigma_\lambda,s_\lambda,\alpha_t)$ defines
in a canonical way the CCR-dynamical system
$\left(\overline{W(M_{2,sa},\sigma_\lambda)},\tilde\omega_\lambda,\tilde\alpha_t\right)$;
$\tilde\omega_\lambda$ is a quasi-free state on the CCR-algebra
$\overline{W(M_{2,sa},\sigma_\lambda)}$:
\begin{align*}
  \tilde\omega_\lambda\bigl(W(A)\bigr)=
  e^{-\frac{1}{2}s_\lambda(A,A)}& &\text{and}&
  &\tilde\alpha_t\bigl(W(A)\bigr) = W\bigl(\alpha_t(A)\bigr)
\end{align*}
for all $A\in M_{2,sa}$.

Let $(\tilde\mathcal{H}_\lambda,\tilde\pi_\lambda,
\tilde\Omega_\lambda)$ be the GNS triplet of $\tilde\omega_\lambda$.
As the state $\tilde\omega_\lambda$ is regular, there exists a real
linear map, called the bose field
$F_\lambda:M_{2,sa}\rightarrow\mathcal{L}(\tilde\mathcal{H}_\lambda)$
such that $\tilde\pi_\lambda\bigl(W(A)\bigr)=e^{iF_\lambda(A)}$ and
the commutation relations $\bigl[F_\lambda(A),F_\lambda(B)\bigr]=
i\sigma_\lambda(A,B)$. As in section \ref{can-coord/norm.fluct.}, a
central limit theorem allows the identification
$\lim_{N\to\infty}F_N(A)=F_\lambda(A)$.  The state
$\tilde\omega_\lambda$ is completely characterized by the two-point
function on the algebra of fluctuations
\begin{equation*}
\tilde\omega_\lambda\bigl( F_\lambda(A)F_\lambda(B) \bigr)
=\lim_{N\to\infty}\omega_\lambda\bigl( F_N(A)F_N(B) \bigr)
= s_\lambda(A,B)+\frac{i}{2}\sigma_\lambda(A,B).
\end{equation*}

Now we proceed to the construction of the complex structure $J$ (see
section \ref{can-coord/can.coord.}). By diagonalisation of the matrix
$h_\lambda$ it is easily seen that $h_\lambda$ has eigenvalues $\pm
\mu$, where $\mu=(\epsilon^2+|\lambda|^2)^{1/2}$.  The spectral
resolution of $h_\lambda$ is hence given by
\begin{equation*}
h_\lambda = -\mu P_- + \mu P_+.
\end{equation*}
In order to construct $J$ we need to know the spectral resolution of
$[h_\lambda,\cdot]$ considered as operator on $M_2$. The spectrum of
$[h_\lambda,\cdot]$ is given by $\{ -2\mu, 0, 2\mu \}$, the
corresponding spectral projections are respectively:
\begin{align*}
  E_- = E(-2\mu)=P_- \cdot P_+ & &E_0= P_-\cdot P_- + P_+\cdot P_+ &
  &E_+= E(2\mu)=P_+ \cdot P_-,
\end{align*}
and $[h_\lambda,A]=-2\mu E_-(A)+2\mu E_+(A)$.

On $M_{2,sa}^1\equiv ( E_+ + E_-)M_{2,sa}$ define $J$ as in section
\ref{can-coord} (equation (\ref{compl-struct})) by
\begin{equation*}
J( E_+ +  E_-)(A) = i( E_+ -  E_-)(A).
\end{equation*}
This operator $J$ is a complex structure on the symplectic space
$(M_{2,sa}^1,\sigma_\lambda)$, satisfying the properties of
Proposition \ref{prop-compl-struct}: $J^2 = -1$, $\sigma_\lambda(A,JB)
= -\sigma_\lambda(JA,B),\;\;\;A,B\in M_{2,sa}^1$ and
$\sigma_\lambda(A,JA)>0\;,\;\;\mathrm{if}\;0\not=A\in M_{2,sa}^1$.
Remark that on $M_{2,sa}^1$, $[h_\lambda,\cdot] = -2i\mu J(\cdot)$
(Cfr. Proposition \ref{pr:norm-mode-A}).

For $\lambda\not=0$, we have $[h_{\lambda},\sigma^z]\not=0$. However
$[h_{\lambda},E_0(\sigma^z)]=0$, and the state $\omega_\lambda$ and
the corresponding time evolution $\alpha_t$ are still invariant under
the symmetry generated by $E_0(\sigma^z)$:
\begin{equation*}
\lim_{N\to\infty}\omega_\lambda\biggl(\Bigl[\sum_{i=-N}^NE_0(\sigma^z)_i,A\Bigr]\biggr)=0
\end{equation*}
for all local $A$. Symmetry breaking is only concerned with the
operator
\begin{equation*}
\hat\sigma^z\equiv\sigma^z-E_0(\sigma^z)=(E_++E_-)(\sigma^z);
\end{equation*}
$\hat\sigma^z\in M_{2,sa}^1$ and we are interested in the fluctuations
of the operator $\hat\sigma^z$ together with its adjoint
$J\hat\sigma^z$.  By calculating $[h_\lambda,\sigma^z]=2\mu(E_+ -
E_-)(\sigma^z)$, we find
\begin{equation*}
J\hat\sigma^z = \frac{i}{\mu}(\lambda\sigma^+ - \bar\lambda \sigma^-).
\end{equation*}
Similarly $[h_\lambda,J\hat\sigma^z]=2i\mu(E_+ + E_-)(\sigma^z)$
yields
\begin{equation*}
\hat\sigma^z =  \frac{|\lambda|^2}{\mu^2}\sigma^z + \frac{\epsilon}{\mu^2}
(\lambda\sigma^+ + \bar\lambda \sigma^-).
\end{equation*}
Note that $J\hat\sigma^z$ is the usual order parameter operator for
the BCS model, but now constructed by means of $\sigma^z$ and the
spectrum of the Hamiltonian.  Therefore it is called the
\emph{canonical order parameter operator}.  We have also
$\omega_\lambda\bigl(J\hat\sigma^z\bigr)=0$ and
$0=\omega_\lambda\bigl([h_\lambda,J\hat\sigma^z]\bigr)=2i\mu\omega_\lambda(\hat\sigma^z)$.

The variances of the fluctuation operators are easily calculated since
\begin{equation*}
(E_0\sigma^z)^2 = \frac{\epsilon^2}{\mu^2} \qquad
(\hat\sigma^z)^2 = (J\hat\sigma^z)^2 = \frac{|\lambda|^2}{\mu^2}.
\end{equation*}
Note $1=(\sigma^z)^2=E_0(\sigma^z)^2+(\hat\sigma^z)^2$.  Also
\begin{equation*}
\rho_\lambda(\sigma^z)=\rho_\lambda(E_0\sigma^z)=-\frac{\epsilon}{\mu}\tanh{\beta\mu}=-2
\epsilon.
\end{equation*}
Hence
\begin{align*}
  \tilde\omega_\lambda \bigl(F_\lambda(E_0\sigma^z)^2\bigr) &=
  s_\lambda(E_0\sigma^z,E_0\sigma^z)
  =\rho_\lambda\bigl((E_0\sigma^z)^2\bigr)
  -\rho_\lambda(E_0\sigma^z)^2 =
  \frac{\epsilon^2}{\mu^2} - 4\epsilon^2\\
  \tilde\omega_\lambda \bigl(F_\lambda(\hat\sigma^z)^2\bigr) &=
  s_\lambda(\hat\sigma^z,\hat\sigma^z)
  =\rho_\lambda\bigl((\hat\sigma^z)^2\bigr)=\frac{|\lambda|^2}{\mu^2}\\
  \tilde\omega_\lambda \bigl(F_\lambda(J\hat\sigma^z)^2\bigr) &=
  s_\lambda(J\hat\sigma^z,
  J\hat\sigma^z)=\rho_\lambda\bigl((J\hat\sigma^z)^2\bigr)=\frac{|\lambda|^2}{\mu^2}.
\end{align*}
The only non-trivial commutator is
\begin{equation*}
\bigl[F_\lambda(\hat\sigma^z),F_\lambda(J\hat\sigma^z)\bigr]=i\sigma_\lambda(\hat\sigma^z,
J\hat\sigma^z)=\omega_\lambda\bigl([\hat\sigma^z,J\hat\sigma^z]\bigr)=
\omega_\lambda\bigl([\sigma^z,J\hat\sigma^z]\bigr)=i\frac{4|\lambda|^2}{\mu},
\end{equation*}
expressing the bosonic character of the fluctuations. Remark on the
other hand that the microscopic observables $\hat\sigma^z$ and
$J\hat\sigma^z$ do not satisfy canonical commutation relations, only
their fluctuations do.

The flucuation operator $F_\lambda(E_0\sigma^z)$ is invariant under
the dynamics $\tilde\alpha_t$, but the operators
$F_\lambda(\hat\sigma^z)$ and $F_\lambda(J\hat\sigma^z)$ satisfy the
equations of motion
\begin{align}
  \frac{d}{idt}\tilde\alpha_t \bigl(F_\lambda(\hat\sigma^z)\bigr)&=
  F_\lambda \bigl([h_\lambda,\alpha_t(\hat\sigma^z)]\bigr)= -2i\mu
  F_\lambda\bigl(\alpha_t(J\hat\sigma^z)\bigr)
  =-2i\mu\tilde\alpha_t F_\lambda(J\hat\sigma^z)\label{macro-dyn1}\\
  \frac{d}{idt}\tilde\alpha_t \bigl(F_\lambda(J\hat\sigma^z)\bigr)&=
  F_\lambda\bigl([h_\lambda,\alpha_t(J\hat\sigma^z)]\bigr)= 2i\mu
  F_\lambda\bigl(\alpha_t(\hat\sigma^z)\bigr) =2i\mu \tilde\alpha_t
  F_\lambda\bigl(\hat\sigma^z\bigr)\label{macro-dyn2}.
\end{align}
In integrated form one gets:
\begin{align*}
  \tilde \alpha_t F_\lambda(\hat\sigma^z) &=
  F_\lambda(\hat\sigma^z)\cos 2\mu t
  + F_\lambda(J\hat\sigma^z) \sin 2\mu t\\
  \tilde \alpha_t F_\lambda(J\hat\sigma^z) &=
  -F_\lambda(\hat\sigma^z)\sin 2\mu t + F_\lambda(J\hat\sigma^z) \cos
  2\mu t.
\end{align*}

Hence by an explicit calculation we have arrived at the results of
Proposition \ref{pr:norm-mode-A}, for $A=\hat \sigma^z$, the generator
of the broken symmetry.  Therefore, denoting $Q_\lambda \equiv
F_\lambda(\hat\sigma^z)$ and $P_\lambda \equiv
\frac{1}{2\mu}F_\lambda(J\hat\sigma^z)$, we defined the pair
$(Q_\lambda,P_\lambda)$ as {\em the canonical pair of the Goldstone
  bosons}. Writing down the previous results in terms of $Q_\lambda$
and $P_\lambda$ (as in Proposition \ref{pr:norm-mode-A}) one sees that
this pair shares indeed all physical properties for Goldstone bosons.

Remark that the frequency of oscillation is $2\mu$.  This is the
phenomenon of the doubling of the frequency for the inherent plasmon
frequency.

The formula
\begin{equation*}
\tilde\omega_\lambda\left( Q_\lambda^2 \right) = (2\mu)^2
\tilde\omega_\lambda\left( P_\lambda^2 \right) =
 \frac{|\lambda|^2}{\mu^2} =\frac{c_\lambda(2\mu)}{2}\coth\frac{\beta(2\mu)}{2},
\end{equation*}
is a quantum mechanical expression of a virial theorem. Remark that in
the normal phase ($\lambda\to 0$), $Q_{\lambda=0}=P_{\lambda=0}=0$,
i.e. the Goldstone boson disappears.

The creation and annihilation operators of the Goldstone bosons are as
usual
\begin{equation*}
a^\pm_\lambda = \frac{ Q_\lambda \mp i 2\mu P_\lambda}{\sqrt{4\mu}}.
\end{equation*}
The state $\tilde\omega_\lambda$ is gauge-invariant and quasi-free
with respect to the gauge transformations of these creation and
annihilation operators, i.e. $\tilde\omega_\lambda\left( a^+_\lambda
  a^+_\lambda \right) = 0 = \tilde\omega_\lambda\left( a^+_\lambda
\right)$, and the two-point function
\begin{equation*}
\tilde\omega_\lambda\left( a^+_\lambda a^-_\lambda \right) = \frac{1}{e^{2\beta\mu}-1}.
\end{equation*}

\section{Short range interactions}\label{short-range}
\subsection{Goldstone theorem and canonical order parameter}
Let $\omega$ be an extremal translation invariant
$(\alpha_t,\beta)$-KMS state, $\alpha_t$ a dynamics generated by a
translation invariant Hamiltonian $H$ and let $\gamma_s$ be a strongly
continuous one-parameter symmetry group which is locally generated by
a generator
\begin{equation*}
Q_n=\sum_{x\in\Lambda_n} q_x,
\end{equation*}
where $\Lambda_n=[-n,n]^\nu \cap \Z^\nu$ and $q_x$ is the symmetry
generator density, i.e.  for $A\in\mathcal{A}_{\Lambda_n}$,
\begin{equation*}
\gamma_s(A)=e^{isQ_n}A\;e^{-isQ_n}.
\end{equation*}
Denote $q=q_{x=0}$, and for convenience denote again $q-\omega(q)$ by $q$.

For systems with short range interactions, assuming spontaneous
symmetry breaking amounts to:
\begin{assumption}\label{assum:ssb}
  Assume that there exists an ($\alpha_t$, $\beta$)-KMS or ground
  state $\omega$ such that $\omega$ is not invariant under the
  symmetry transformation $\gamma$, while the dynamics $\alpha_t$
  remains invariant under $\gamma$, i.e. 
  \begin{align}
    &\exists A\in\mathcal{A}_L\;\;\text{such that}\;\;\omega\bigl(\gamma_s(A)
    \bigr)\not=\omega(A)\label{eq:state-ssb}\\
    &\alpha_t \circ \gamma_s = \gamma_s \circ \alpha_t.\label{eq:t-invar}
  \end{align}
\hspace*{\fill}\qed
\end{assumption}

The invariance of the dynamics (\ref{eq:t-invar}) is crucial in this
context (see \cite{requardt:1982} and Proposition \ref{pr:ord-par} and
equation (\ref{eq:gapless}) below).  For a more complete discussion of
the phenomenon of spontaneous symmetry breaking, see
\cite{wreszinski:1987}.

An operator $A$ satisfying (\ref{eq:state-ssb}) is called an order
parameter operator. Eq. (\ref{eq:state-ssb}) is equivalent to
\begin{equation*}
\frac{d}{ds}\Bigl.\omega\bigl(\gamma_s(A)\bigr)\Bigr|_{s=0}= 
\lim_{n\to\infty}\omega\bigl( [Q_n,A] \bigr) \not=0.
\end{equation*}

The local Hamiltonians are determined by an interaction $\Phi$
$H_n = \sum_{X\subseteq \Lambda_n}\Phi(X)$
and the infinite volume Hamiltonian $H$ is defined such that for $A
\in \mathcal{A}_ {\Lambda_0}$,
\begin{equation*}
H A\Omega= \sum_{X\cap\Lambda_0\not=\varnothing}[\Phi(X),A]\Omega,
\end{equation*}
where $\Omega$ is the cyclic vector of the state $\omega$.

The relation between spontaneous symmetry breaking and the absence of
a gap in the energy spectrum in the ground state was originally put
forward by Goldstone \cite{goldstone:1961}. For short range
interactions in many-body systems, it is proved
\cite{kastler:1966,swieca:1967} that spontaneous symmetry breaking
implies the absence of an energy gap in the excitation spectrum. We
refer here to \cite{landau:1981} where the Goldstone theorem is proved
rigorously for quantum lattice systems.
\begin{theorem}[Goldstone Theorem \cite{landau:1981}]
  If $\Phi$ is translation invariant and satisfies
\begin{equation}\label{sh-rg}
\sum_{X\ni 0} |X| \|\Phi(X)\| <\infty
\end{equation}
then
\begin{enumerate}
\item At $T=0$: If the system has an energy gap then there is no
  spontaneous symmetry breakdown.
  
\item At $T>0$: If the system has $L^1$ clustering then there is no
  spontaneous symmetry breakdown.\qed
\end{enumerate}
\end{theorem}
The $L^1$ clustering means here that for each observable $A$, one has:
\begin{equation*}
\sum_{x\in\Z^\nu}\bigl|\omega(A\tau_xA)-\omega(A)^2\bigr|<\infty.
\end{equation*}

The first step is to construct something like a {\em canonical order
  parameter operator}. See section \ref{long-range} for an example of
this construction.  Denote
\begin{equation*}
L(A)=[H,A].
\end{equation*}
The Duhamel two-point function becomes now:
\begin{equation*}
(A,B)_\sim \equiv \frac{1}{\beta}\int_0^\beta \omega(A^*\alpha_{iu}B) du =
\omega\left( A^*\frac{1-e^{-\beta L}}{\beta L} B\right).
\end{equation*}
The KMS-condition,
$\omega\left( AB \right) = \omega\left(B\alpha_{i\beta}A\right)$,
yields
\begin{equation*}
\omega\bigl([A,B]\bigr)=\omega\Bigl( A(1-e^{-\beta L})B \Bigr),
\end{equation*}
for $A,B$ in a dense domain of $\mathcal{A}$, and hence if $B\in
\mathrm{Dom}(L^{-1})$ then
\begin{equation*}
\beta (A,B)_\sim = \omega\Bigl(\bigl[A,L^{-1}B\bigr]\Bigr).
\end{equation*}
and the Bogoliubov inequality
\cite{bogoliubov:1962} for KMS-states is given by:
\begin{equation*}
\bigl| \omega\bigl( [A^*,B] \bigr)\bigr|^2 \leq \beta
 \omega\bigl( [ A^*, L(A) ] \bigr) (B,B)_\sim.
\end{equation*}
Finally denote the local $0$-mode fluctuation of an observable $A$ in
the state $\omega$ by
\begin{equation*}
F_{n,0}(A)=\frac{1}{|\Lambda_n|^{1/2}}\sum_{x\in\Lambda_n}\tau_x  A - \omega(A).
\end{equation*}
\begin{assumption}\label{assum:q-fluc-finite}
  Assume that there are no long range correlations in the fluctuations
  of the symmetry generator density, i.e. assume
\begin{equation*}
\lim_{n\to\infty}\omega\bigl(F_{n,0}(q)^2\bigr)=
\sum_{z\in\Z^\nu}\left|\omega(q\tau_z q)-\omega(q)^2\right| <\infty.
\end{equation*}
\hspace*{\fill}\qed
\end{assumption}
Then also the uniform susceptibility $c_0^\beta$ defined by
\begin{equation}\label{eq:c_0^beta}
c_0^\beta\equiv \lim_{n\to\infty}\frac{\beta}{2}\bigl(F_{n,0}(q),F_{n,0}(q) \bigr)_\sim
\end{equation}
is finite, i.e. $c_0^\beta <\infty$.

\begin{proposition}\label{pr:ord-par}
  Under Assumption \ref{assum:ssb} and \ref{assum:q-fluc-finite} we
  have
\begin{equation}\label{c-beta}
c_0^\beta = \lim_{n\to\infty}\frac{\beta}{2}
\bigl(F_{n,0}(q),\alpha_t F_{n,0}(q)\bigr)_\sim >0
\end{equation}
and $c_0^\beta$ is independent of $t$, and  given by
\begin{equation*}
c_0^\beta=\lim_{n\to\infty}\frac{1}{2}\omega\Bigl( \bigl[Q_{n},
L^{-1}(q)\bigr]\Bigr).
\end{equation*}
\end{proposition}
\begin{proof}
  Let
\begin{equation*}
c_0^\beta(t)=
\lim_{n\to\infty}\frac{\beta}{2}
\bigl(F_{n,0}(q),\alpha_t
F_{n,0}(q)\bigr)_\sim=\lim_{n\to\infty}\frac{1}{2}
\omega\Bigl( F_{n,0}(q)\frac{1-e^{-\beta L}}{L} e^{itL} F_{n,0}(q)\Bigr).
\end{equation*}
First we show $c_0^\beta(t=0)>0$. Let $A$ be an arbitrary order
parameter operator.  SSB, translation invariance and the Bogoliubov
inequality yield
\begin{align*}
  0&< \lim_{n\to\infty} \Bigl|\omega\bigl(
  \bigl[F_{n,0}(q),F_{n,0}(A)\bigr] \bigr)
  \Bigr|^2 \\
  &\leq \lim_{n\to\infty} \beta \omega\Bigl( \bigl[F_{n,0}(A),
  L\bigl(F_{n,0}(A)\bigr) \bigr] \Bigr) \bigl(F_{n,0}(q),
  F_{n,0}(q)\bigr)_\sim.
\end{align*}
In \cite{landau:1981} it is shown that (\ref{sh-rg}) also implies
\begin{equation*}
\lim_{n\to\infty}\omega\Bigl( \bigl[F_{n,0}(A), L\bigl(F_{n,0}(A)\bigr) \bigr] \Bigr)
=\sum_{z\in\Z^\nu}\omega \Bigl(\bigl[\tau_z A, L(A)\bigr]\Bigr)<\infty
\end{equation*}
for each local observable $A$.  Hence
\begin{equation*}
0<\sum_{z\in\Z^\nu}\omega \Bigl(\bigl[\tau_z A, L(A)\bigr]\Bigr)\lim_{n\to\infty}
\beta \bigl(F_{n,0}(q), F_{n,0}(q)\bigr)_\sim
\end{equation*}
yielding $c_0^\beta(t=0)>0$.

The proof of the time invariance of $c_0^\beta$ is based on
\cite{requardt:1982} and goes as follows:
\begin{align*}
  \frac{d}{idt}c_0^\beta(t) &= \lim_{n\to\infty}\frac{\beta}{2}
  \Bigl(F_{n,0}(q),\alpha_t L\bigl(F_{n,0}(q)\bigr)\Bigr)_\sim =
  \lim_{n\to\infty}\frac{1}{2}
  \omega\Bigl( F_{n,0}(q)(1-e^{-\beta L}) e^{itL} F_{n,0}(q)\Bigr)\\
  &= \lim_{n\to\infty}\frac{1}{2} \omega\Bigl(
  \bigl[F_{n,0}(q),e^{itL}F_{n,0}(q)\bigr]\Bigr).
\end{align*}
Translation invariance and (\ref{eq:t-invar}) yield:
\begin{align*}
  \frac{d}{idt}c_0^\beta(t)
  &=\lim_{n\to\infty}\frac{1}{2}\omega\Bigl(\bigl[Q_n,e^{itL}q\bigr]\Bigr)
  =\frac{1}{2}\left.\frac{d}{ids}\right|_{s=0}
  \omega\bigl(\gamma_s(\alpha_t q)\bigr)
  =\frac{1}{2}\left.\frac{d}{ids}\right|_{s=0} \omega\bigl(\alpha_t(\gamma_s  q)\bigr)\\
  &=\frac{1}{2}\left.\frac{d}{ids}\right|_{s=0} \omega( q) = 0.
\end{align*}

\end{proof}

From the proposition it follows that if $L^{-1} q$ exists, it is an
order parameter operator. We call it the \emph{canonical order
  parameter operator}, it is an order parameter constructed directly
from the two given quantities, the Hamiltonian and the symmetry
generator.  However it can not be expected in general that
$q\in\mathrm{Dom}(L^{-1})$, especially not for systems without an
energy gap, because of problems at zero energy.  Expressions like
$(1-e^{-\beta L})L^{-1} q$ on the contrary are well defined. The bulk
of our efforts below consists of mastering the difficulties with the
canonical order parameter by considering the $k$-mode fluctuations and
by afterwards taking the limit $k\to 0$. This method has already been
used in \cite{michoel:1999b}, where the Goldstone coordinates are
constructed for models of interacting Bose gases.

\subsection{Fluctuations}
By the Goldstone theorem, spontaneous symmetry breaking implies that
the system does not have exponential or $L^1$ clustering. In
particular the variances of local fluctuations $F_{n,0}(A)$ may not be
convergent in the thermodynamic limit for certain $A$ (in particular
for $A$ an order parameter operator) because of long range order
correlations.  The central limit as described in section
\ref{can-coord/norm.fluct.}  no longer holds. However one can study
the $k$-mode fluctuations, i.e. one considers for $k=
(k_1,k_2,\ldots,k_\nu)\in \R^\nu$, with $k_j\not=0$ for
$j=1,2,\ldots,\nu$:
\begin{equation*}
F_{n,k}(A)=\frac{1}{|\Lambda_n|^{1/2}}\sum_{x\in\Lambda_n}\bigl(\tau_x(A)-\omega(A)\bigr)
\cos k.x.
\end{equation*}

It is believed that the central limit theorem holds for the $k$-mode
fluctuations in every extremal translation invariant state, even at
criticality. This is essentially because one stays away from the
singularity at $k=0$. A completely rigorous proof of this statement is
found in \cite{michoel:1998}, for the absolute convergent case under a
very mild cluster condition. Below we prove the convergence of the
Fourier series for translation invariant states with singularities
occuring only at zero momentum (see further on). See also
\cite{narnhofer:1983} for a similar line of reasoning.

For $A\in\mathcal{A}_L$, denote the Fourier transforms of the
$l$-point correlation functions $\omega(\tau_{x_1}A\tau_{x_2}A\cdots
\tau_{x_l}A)$ by $\mu(k_1,k_2,\cdots,k_l)$ (i.e. $k_j$ are different
vectors in $\R^\nu$ here, not the components of a particular $k$).  In
general $\mu$ is a measure. By translation invariance it can be
written as a function of $k_1,k_1+k_2,\ldots,k_1+k_2+\cdots+k_l$.  As
in \cite{narnhofer:1983}, assume that the only singularities in $\mu$
are of the type $\delta(k_1+\cdots+k_i)$ (i.e. singularities occuring
only at zero momentum).

We show now that the truncated correlation functions $\omega_T\left(
  F_{n,k}(A)^l\right)$ vanish for $l\geq 3$ and remain finite for
$l=2$. Let $\omega(A)=0$, then
\begin{align*}
  &\omega_T\left( F_{n,k}(A)^l\right)\\
  &=\frac{1}{|\Lambda_n|^{l/2}}\sum_{x_1,x_2,\ldots,x_l}\omega_T(\tau_{x_1}A\tau_{x_2}A\cdots
  \tau_{x_l}A)\cos k.x_1 \cos k.x_2 \cdots \cos k.x_l\\
  &=\frac{1}{|\Lambda_n|^{l/2}}\sum_{x_1,y_1\ldots,y_{l-1}}\omega_T(A\tau_{y_1}A\cdots
  \tau_{y_{l-1}}A)\cos k.x_1 \cos k.(y_1+x_1)\cdots \cos
  k.(y_{l-1}+x_1).
\end{align*}
The expansion of the cosines into exponentials yields two types of
terms, namely terms which do not depend on $x_1$ and terms which do
depend on $x_1$. The first kind of terms do not appear for $l$ odd and
for $l$ even they are exactly the ones which are cancelled out by the
truncation. The second kind of terms tend to zero because of the
scaling factors. Let us illustrate this by means of an example. First
let $l=2$:
\begin{equation*}
\omega_T\left( F_{n,k}(A)^2 \right)=\omega\left( F_{n,k}(A)^2 \right)
=\frac{1}{|\Lambda_n|}\sum_{x,y}\omega(A \tau_{y-x} A)\cos k.x \cos k.y.
\end{equation*}
Since
\begin{equation*}
\mu(k) =\sum_z \omega(A \tau_{z} A) e^{-ik.z}
\end{equation*}
can at most have a singularity at $k=0$, $\mu(k)<\infty$ for
$k\not=0$. Also
\begin{equation*}
\frac{1}{|\Lambda_n|}\sum_z e^{ik.z}\to \delta_{k,0}.
\end{equation*}
Hence the only terms contributing in the two-point correlation
function are the terms containing the factor $e^{\pm ik.(y-x)}$, i.e.
the terms of the first kind. In the limit we find
\begin{equation*}
\lim_n \omega_T\left( F_{n,k}(A)^2 \right) = \frac{1}{4}[\mu(k)+\mu(-k)]<\infty.
\end{equation*}

Now let $l=4$ and consider a typical term:
\begin{equation*}
\frac{1}{|\Lambda_n|^2}\sum_{x_1,y_1,y_2,y_3}\omega\left( A\tau_{y_1}A\tau_{y_2}A
\tau_{y_3}A \right)e^{-ik.(y_1-y_2+y_3)}.
\end{equation*}
Ignoring boundary effects in the sums, this becomes
\begin{align*}
  &\frac{1}{|\Lambda_n|}\sum_{y_1,y_2,y_3}\omega\left(
    A\tau_{y_1}A\tau_{y_2}A
    \tau_{y_3}A \right)e^{-ik.(y_1-y_2+y_3)}\\
  &=\frac{1}{|\Lambda_n|}\sum_{y_1,y_2,y_3}\omega\left(
    A\tau_{y_1}A\tau_{y_2}[A\tau_{y_3-
      y_2}A]\right)e^{-ik.(y_1-y_2+y_3)}\\
  &=\sum_{x,z}\omega\Bigl(
  A\tau_{x}A\frac{1}{|\Lambda_n|}\sum_y\tau_{y}[A\tau_zA]\Bigr)
  e^{-ik.(x+z)}.
\end{align*}
In the limit we get
\begin{equation*}
\sum_x \omega\left( A\tau_{x}A\right)e^{-ik.x}\sum_z \omega\left( A\tau_{z}A\right)
e^{-ik.z}
\end{equation*} 
cancelling out against two-point correlations in the 4-point truncated
correlation function.

Finally, take $l=3$, then all terms are of the second kind and vanish,
e.g.
\begin{align*}
  &\frac{1}{|\Lambda_n|^{3/2}}\sum_{x_1,y_1,y_2}\omega(A\tau_{y_1}A\tau_{y_2}A)
  e^{ik.(x_1+y_1-y_2)}\\
  =&\frac{1}{|\Lambda_n|^{3/2}}\sum_{y_1,y_2}\omega(A\tau_{y_1}[A\tau_{y_2-y_1}A])
  e^{ik.(y_1-y_2)}\sum_{x_1}e^{ik.x_1}.
\end{align*}
The sum over $x_1$ is bounded by $\prod_{j=1}^\nu |\sin
\frac{k_j}{2}|^{-1}$, yielding
\begin{equation*}
\sum_y\omega\Bigl(A\frac{1}{|\Lambda_n|}\sum_x\tau_x[A\tau_yA]\Bigr)e^{-ik.y}
\end{equation*}
which converges to
\begin{equation*}
\omega(A)\sum_y \omega\left( A\tau_{y}A\right)e^{-ik.y}=0.
\end{equation*}

Using the formula
\begin{equation*}
\omega\left( e^{i\lambda Q} \right) = \exp\biggl\{ \sum_{l=1}^\infty\frac{(i\lambda)^l}
{l!}\omega_T\Bigl( \underbrace{Q,\ldots,Q}_{l\; times} \Bigr) \biggr\}
\end{equation*}
one arrives at the central limit theorem
\begin{equation*}
\lim_n\omega\left(e^{iF_{n,k}(A)}\right) = e^{-\frac{1}{2}s_k(A,A)},
\end{equation*}
with $s_k(A,A)=\lim_{n\to\infty}\omega\left(F_{n,k}(A)^2\right)$.

In \cite{michoel:1998} one can find a rigorous proof of the central
limit theorem for the $k$-mode fluctuations, $k=(k_j\not=
0)_{j=1}^\nu$, for states satisfying a certain clustering condition,
expressed as a condition on the function $\alpha_\omega$ (see equation
(\ref{cluster-f})).  Although this condition is much weaker than for
the $k=0$ fluctuations, it is not clear whether it is always satisfied
for any extremal translation invariant state.  The arguments above
however suggests that this clustering condition on the state is merely
technical and that a general rigorous proof of the central limit
theorem along the lines of \cite{michoel:1998} is possible for
$k=(k_j\not= 0)_{j=1}^\nu$ under even weaker conditions. We continue
on the basis of the arguments above.

\begin{theorem}[Central limit theorem]\label{CLT-k}
  If the state $\omega$ has only singularities at zero momentum, for
  all $A\in\mathcal{A}_{L,sa}$ and $k=(k_j\not= 0)_{j=1}^\nu$, then
\begin{enumerate}
\item $\lim_{n\to\infty}\omega\left(F_{n,k}(A)^2\right) <\infty$
  
\item $ \lim_{n\to\infty}\omega\left(e^{iF_{n,k}(A)}\right) =
  e^{-\frac{1}{2}s_k(A,A)}$ with
  $s_k(A,B)=\lim_{n\to\infty}\mathrm{Re}\;\omega\left(F_{n,k}(A)^*
    F_{n,k}(B)\right)$.\qed
\end{enumerate}
\end{theorem}

Because of $(i)$, the limit
\begin{equation*}
\lim_{n\to\infty}\omega\bigl(F_{n,k}(A)^*F_{n,k}(B)\bigr)\equiv \langle A,B \rangle_k
\end{equation*}
defines a positive sesquilinear form which satisfies the
Cauchy-Schwarz inequality
\begin{equation*}
|\langle A,B \rangle_k|^2 \leq \langle A,A \rangle_k \langle B,B \rangle_k.
\end{equation*}
More explicitly
\begin{equation*}
\langle A,B \rangle_k = \frac{1}{2}\sum_{z\in\Z^\nu}\bigl( \omega(A^*\tau_z B)
-\omega(A^*)\omega(B)\bigr)\cos k.z.
\end{equation*}
Let
\begin{equation*}
\sigma_k(A,B)=2\;\mathrm{Im}\;\langle A,B \rangle_k, 
\end{equation*}
then
\begin{equation*}
\mathrm{strong}-\lim_{n\to\infty}\pi\bigl( [F_{n,k}(A),F_{n,k}(B)] \bigr) =
i\sigma_k(A,B).
\end{equation*}

The identification of the central limit with bose fields is as in
section \ref{can-coord/norm.fluct.}, and worked out in full detail for
$k\not= 0$ in \cite{michoel:1998}. The bilinear form $s_k$ determines
a quasi free state $\tilde\omega_k$ on the CCR-algebra
$\mathcal{W}(\mathcal{A}_{L,sa},\sigma_k)$:
\begin{equation*}
\tilde\omega_k \left( W_k(A) \right) = e^{-\frac{1}{2}s_k(A,A)}.
\end{equation*}
The $W_k(A)$, $A\in\mathcal{A}_{L,sa}$ are the Weyl operators
generating $\mathcal{W}(\mathcal{A}_{L,sa},\sigma_k)$.  Via the
central limit theorem, one shows for $ A_1 , A_2 ,\ldots , A_l \in
\mathcal{A}_{L,sa} $,
\begin{equation*}
\lim_{n \to\infty}\omega\Bigl( e^{i F_{n,k} (A_1)} e^{i F_{n,k} (A_2)}\ldots
e^{i F_{n,k} (A_l)}\Bigr) = \tilde\omega_k\Bigl( W_k(A_1) W_k(A_2) \ldots W_k(A_l)\Bigr).
\end{equation*}
The state $ \tilde\omega_k $ is regular and hence for every $A \in
\mathcal{A}_{L,sa}$ there exists a self-adjoint bosonic field $F_k(A)$
in the GNS representation
$(\tilde\mathcal{H}_k,\tilde\pi_k,\tilde\Omega_k)$ of $\tilde\omega_k$
such that
\begin{equation*}
\tilde\pi_k ( W_k (A) ) = e^{iF_k(A)}.
\end{equation*}
This implies that in the sense of the central limit, the local
fluctuations converge to the bosonic fields associated with the system
$\Bigl({\cal W} (\mathcal{A}_{L,sa} , \sigma_k ),\tilde\omega_k
\Bigr)$,
\begin{equation*}
\lim_{n \to \infty } F_{n,k} (A)= F_k(A).
\end{equation*}
As in section \ref{can-coord/norm.fluct.}, fluctuation operators are
only defined up to equivalence i.e. $A\equiv_k B$ if\\ $\langle
A-B,A-B\rangle_k=0$ and
\begin{equation}\label{eq:equiv-k}
  A\equiv_k B \Leftrightarrow \tilde \pi_k\bigl(W_k(A)\bigr)= \tilde\pi_k
  \bigl(W_k(B)\bigr).
\end{equation}
The form $\langle\cdot,\cdot\rangle_k$ thus becomes a scalar product
on $[\mathcal{A}_L]$, the equivalence classes of $\mathcal{A}_L$ for
the relation $\equiv_k$. Denote by $\mathcal{K}_k$ the Hilbert space
obtained as completion of $[\mathcal{A}_L]$ and by
$\mathcal{K}_k^{Re}$ the real subspace of $\mathcal{K}_k$ generated by
$[\mathcal{A}_{L,sa}]$.

\subsection{Goldstone modes for finite wavelengths}\label{knot0}
The finiteness of $\lim_{n\to\infty}\omega\bigl(F_{n,k}(q)^2\bigr)$
for all $k$ ($k=0$ included by Assumption \ref{assum:q-fluc-finite})
implies the finiteness of
\begin{equation*}
\lim_{n\to\infty}\int |\hat f(\lambda)|\omega\bigl(F_{n,k}(q)dE_\lambda F_{n,k}(q)\bigr)
\end{equation*}
for $\hat f \in\mathcal{D}$, and hence the existence of a measure
\begin{equation*}
d\tilde\mu_k(\lambda)=\lim_{n\to\infty}\omega\bigl(F_{n,k}(q)dE_\lambda F_{n,k}(q)\bigr);
\end{equation*}
$dE_\lambda$ is the spectral measure of the Hamiltonian $H$, i.e.
$H=\int \lambda dE_\lambda$.

As in section \ref{can-coord/2.5}, define the measure
$dc_k^\beta(\lambda)$ with support on $\R^+$ only by
\begin{equation*}
dc_k^\beta(\lambda) = 2\frac{1-e^{-\beta\lambda}}{\lambda}d\tilde\mu_k(\lambda),
\end{equation*}
such that for $\hat f \in\mathcal{D}$ (cfr. Lemma \ref{le:f-correl})
\begin{equation}\label{eq:k-correl-f}
\lim_{n\to\infty}\int \hat f(\lambda)\omega\bigl(F_{n,k}(q)dE_\lambda F_{n,k}(q)\bigr)
=\int_0^\infty \bigl( \hat f(\lambda)+\hat f(-\lambda)e^{-\beta\lambda}\bigr)
\frac{\lambda}{2(1-e^{-\beta\lambda})}dc_k^\beta(\lambda).
\end{equation}
\begin{proposition}\label{pr:0-delta}
  For $\hat f \in\mathcal{D}$,
\begin{equation*}
\lim_{k\to 0}\int_0^\infty \hat f(\lambda)dc_k^\beta(\lambda) 
=\int_0^\infty \hat f(\lambda)dc_0^\beta(\lambda)
= c_0^\beta \hat f(0),
\end{equation*}
where $c_0^\beta$ is given by equation (\ref{eq:c_0^beta}).  In other
words $\lim_{k\to 0}dc_k^\beta(\lambda)=dc_0^\beta(\lambda)=
c_0^\beta\delta(\lambda)d\lambda$.
\end{proposition}
\begin{proof}
  The statement that $\lim_{k\to
    0}dc_k^\beta(\lambda)=dc_0^\beta(\lambda)$ follows from Assumption
  \ref{assum:q-fluc-finite}.  The proof of the second statement is
  based on the time invariance of $c_0^\beta(t) = \lim_{n\to\infty}
  \beta\bigl( F_{n,0}(q),\alpha_t F_{n,0}(q) \bigr)_\sim$ (Proposition
  \ref{pr:ord-par}) and by (\ref{eq:k-correl-f}): for $\hat f
  \in\mathcal{D}$,
\begin{align*}
  \hat f(\lambda)c_0^\beta &= \beta\lim_{n\to\infty}\int f(t)\bigl(
  F_{n,0}(q),\alpha_t
  F_{n,0}(q) \bigr)_\sim e^{-i\lambda t}dt\\
  &=\int_0^\infty \hat f(\lambda-\lambda')dc_0^\beta(d\lambda')
\end{align*}
i.e. $dc_0^\beta(\lambda)=c_0^\beta\delta(\lambda)d\lambda$.  
\end{proof}
In order not to obscure the construction of the Goldstone boson normal
coordinates by technical details, we will first consider the
case that
\begin{equation}\label{eq:approx-delta}
dc_k^\beta(\lambda)=c_k^\beta\delta(\lambda-\epsilon_k^\beta)d\lambda,
\end{equation}
with $\epsilon_k^\beta>0$ and $c_k^\beta=\lim_{n\to\infty} \beta\bigl(
F_{n,k}(q),\alpha_t F_{n,k}(q) \bigr)_\sim$.  From Proposition
\ref{pr:0-delta} we deduce that this is a good approximation for
sufficiently small $|k|$, and we will show later that this approximation
becomes exact in a certain limit $k\to 0$, to be specified later.

From equation (\ref{eq:k-correl-f}) and (\ref{eq:approx-delta}), it
follows
\begin{equation}\label{k-correl-f-delta}
\lim_{n\to\infty}\omega\left( F_{n,k}(q) \hat f(H) F_{n,k}(q) \right)
= \frac{c_k^\beta \epsilon_k^\beta}{2(1-e^{-\beta\epsilon^\beta_k})}
\Bigl( \hat f(\epsilon^\beta_k) + \hat f(-\epsilon^\beta_k) e^{-\beta\epsilon^\beta_k}
\Bigr).
\end{equation}
In particular one has
\begin{equation*}
\tilde\omega_k\left( F_k(q)^2 \right) = \lim_{n\to\infty}\omega\left( F_{n,k}(q)^2 \right)
=\frac{c^\beta_k \epsilon^\beta_k}{2}\coth\frac{\beta\epsilon^\beta_k}{2}.
\end{equation*}
Also time invariance of $c_0^\beta(t)$ (see above) (i.e. SSB) implies
\begin{equation}\label{eq:gapless}
\lim_{k\to 0}\epsilon^\beta_k = 0,
\end{equation}
as can be seen from (\ref{k-correl-f-delta}):
\begin{equation*}
c_0^\beta(t)=\lim_{k\to 0}c_k^\beta(t)=\lim_{k\to 0}c_k^\beta \cos\epsilon^\beta_k t.
\end{equation*}

For $\hat f\in\mathcal{D}$, denote 
\begin{equation*}
  q(f)=\int f(t)\alpha_{-t} q=\hat f(L)q
\end{equation*}
and consider the equivalence class $[q(f)]_k$. For $q(f)\in\mathcal{A}_{L,sa}$ the fluctuation operator $F_k\bigl([q(f)]_k\bigr)$ is well defined,
\begin{equation}\label{eq:F_k-f}
  \tilde\omega_k\Bigl(F_k\bigl([q(f)]_k\bigr)^2\Bigr) 
  = \bigl\langle [q(f)]_k,[q(f)]_k \bigr\rangle_k
  = |\hat f(\epsilon_k^\beta)|^2\frac{c^\beta_k \epsilon^\beta_k}{2}
  \coth\frac{\beta\epsilon^\beta_k}{2},
\end{equation}
(we used that $q(f)\in\mathcal{A}_{L,sa}$
iff$\Bar{\Hat{f}}(\lambda)=\hat f(-\lambda)$ ), and obviously for these
functions $f$, we can define elements $[q]_k(f)\in \mathcal{K}_k^{Re}$
through the relation $[q]_k(f)=[q(f)]_k$. However since
$\mathcal{K}_k$ is by definition closed for the
$\langle\cdot,\cdot\rangle_k$ topology, we can define elements
$[q]_k(f)$ for a much wider class of functions $\mathcal{F}$, namely all those
functions for which $|\hat f(\epsilon_k^\beta)|<\infty$: let $f_i$ be a
sequence of functions such that $[q(f_i)]_k\in \mathcal{K}_k^{Re}$ and
$\lim_i \hat f_i(\epsilon_k^\beta)=\hat f(\epsilon_k^\beta)$,
and define
\begin{equation*}
  [q]_k(f)= \text{strong-}\lim_i[q(f_i)]_k.
\end{equation*}
In particular we have 
\begin{equation*}
  [q]_k(g)\in\mathcal{K}_k^{Re} \;\text{with}\;\hat g(\lambda)=
  \frac{i}{\lambda},
\end{equation*}
and obviously we interpret $F_k\bigl([q]_k(g)\bigr)$ as
``$F_k\bigl(iL^{-1}(q)\bigr)$'', i.e. as the $k$-fluctuation operator
of the canonical order parameter, even though $iL^{-1}(q)$ does not
exist in general.

In the spirit of Proposition \ref{pr:norm-mode-A}, denote
\begin{align*}
  Q_k = F_k(q) && P_k = F_k\bigl([q]_k(g)\bigr)\;\text{with}\;\hat g(\lambda)=
  \frac{i}{\lambda},
\end{align*}
and denote by $\tilde\mathcal{B}_k$ the algebra generated by $Q_k$ and
$P_k$. Also denote by $\tilde\mathcal{C}_k$ the algebra generated by
the operators $F_k\bigl([q]_k(f)\bigr)$ with $f\in\mathcal{F}$.  Our
main result is then that the pair $(Q_k,P_k)$, constructed directly
from the generator of the broken symmetry, forms a harmonic normal
mode, therefore properly called the \emph{Goldstone boson normal
  mode}. This result is an extension of Proposition
\ref{pr:norm-mode-A} to the case of $k\not=0$ fluctuations in the
presence of SSB.

\begin{theorem}\label{thm:gold-boson}
  In the presence of SSB (Assumption \ref{assum:ssb}), and in the case
   (\ref{eq:approx-delta}), the generator of the broken
  symmetry determines uniquely the construction of a canonical pair of
  fluctuation operators $(Q_k,P_k)$,
  \begin{equation*}
    [Q_k,P_k] = ic^\beta_k
  \end{equation*}
  with $c^\beta_k =\lim_{n\to\infty}\beta\Bigl( F_{n,k}(q),F_{n,k}(q)
  \Bigr)_\sim>0$,  satisfying a \emph{virial
    theorem}:
\begin{equation*}
  \tilde\omega_k\left( Q_k^2 \right) = (\epsilon^\beta_k)^2
  \tilde\omega_k\left( P_k^2 \right).
\end{equation*}
The microscopic time evolution $\alpha_t$ induces a time evolution
$\tilde\alpha_t^k$ on $\tilde\mathcal{C}_k$ through the relation
\begin{equation*}
  \tilde\alpha_t^k F_k\bigl([q]_k(f)\bigr) \equiv
  F_k\bigl([q]_k(U_t f)\bigr)\; , \;\; (\widehat{U_t f})(\lambda)=e^{it\lambda}
  \hat f(\lambda);
\end{equation*}
$\tilde\alpha_t^k$ leaves $\tilde\mathcal{B}_k$ invariant and leads to
the equations of motion
\begin{align}
 \tilde\alpha_t^k Q_k &= Q_k \cos\epsilon^\beta_k t +
  \epsilon^\beta_k P_k \sin\epsilon^\beta_k t \label{eq:3} \\
 \tilde\alpha_t^k P_k &=
  -\frac{Q_k}{\epsilon^\beta_k}\sin\epsilon^\beta_k t + P_k
  \cos\epsilon^\beta_k t.\label{eq:4}
\end{align}
The operators $(Q_k,P_k)$ are called the \emph{Goldstone boson normal
  coordinates}. The Goldstone boson creation and annihilation
operators are defined by
\begin{equation*}
  a^\pm_k = \frac{Q_k \mp i\epsilon^\beta_kP_k}{\sqrt{2\epsilon^\beta_k}}
 \end{equation*}
 satisfying $[a^-_k,a^+_k]=c^\beta_k$.  The quasi-free state
 $\tilde\omega_k$ is a $\beta$-KMS state on $\tilde\mathcal{B}_k$ for
 the evolution $\tilde\alpha_t^k$, i.e.  the Goldstone bosons have a
 Bose-Einstein distribution:
\begin{equation*}
\tilde\omega_k\left( a^+_ka^-_k \right) = \frac{c^\beta_k}{e^{\beta\epsilon^\beta_k}-1},
\end{equation*}
which is equivalent to
\begin{equation*}
\tilde\omega_k\left( Q_k^2 \right) = \frac{c^\beta_k
\epsilon^\beta_k}{2}\coth\frac{\beta\epsilon^\beta_k}{2}.
\end{equation*}
The state $\tilde\omega_k$ is gauge invariant: $\tilde\omega_k\left(
  a^+_ka^+_k \right) = 0 = \tilde\omega_k\left( a^+_k \right)$.
\end{theorem}
\begin{proof}
The commutator follows from 
\begin{equation*}
  \sigma_k\bigl([q]_k,[q]_k(g)\bigr)=-i\int\hat g(\lambda)(1-e^{-\beta\lambda})
  d\tilde\mu_k(\lambda).
\end{equation*}
The variance of $P_k$ is obtained from (\ref{eq:F_k-f}):
\begin{equation*}
  \tilde\omega_k\bigl(P_k^2\bigr)=\frac{c_k^\beta}{2\epsilon_k^\beta}\coth
  \frac{\beta\epsilon_k^\beta}{2}=\frac{1}{(\epsilon_k^\beta)^2}
  \tilde\omega_k\bigl(Q_k^2\bigr).
\end{equation*}
Denote $\hat h(\lambda)=i\lambda$. Clearly the infinitesimal generator
of $\tilde\alpha_t^k$ is given by
\begin{equation*}
  \bigl.\frac{d}{dt}\tilde\alpha_t^k\bigr|_{t=0}F_k\bigl([q]_k(f)\bigr)
  =F_k\bigl([q]_k(hf)\bigr).
\end{equation*}
Hence the first relation
\begin{equation}\label{eq:1}
  \Bigl.\frac{d}{dt}\tilde\alpha_t^k\Bigr|_{t=0}P_k = -Q_k
\end{equation}
follows trivially. The second,
\begin{equation}\label{eq:2}
  \Bigl.\frac{d}{dt}\tilde\alpha_t^k\Bigr|_{t=0}Q_k = (\epsilon_k^\beta)^2 P_k,
\end{equation}
follows from the equivalence relation (\ref{eq:equiv-k}): from
equation (\ref{k-correl-f-delta}) one computes straightforwardly
\begin{equation*}
  \bigl\langle [q]_k(h)-(\epsilon_k^\beta)^2[q]_k(g) ,
   [q]_k(h)-(\epsilon_k^\beta)^2[q]_k(g)\bigr\rangle_k =0,
\end{equation*}
where $\hat g(\lambda)=i\lambda^{-1}$ as before.
Exponentiation of (\ref{eq:1}) and (\ref{eq:2}) leads to the equations
of motion. Also the remainder of the theorem follows from
(\ref{k-correl-f-delta}). 
\end{proof}
Remark that for $k\to 0$, $\tilde\omega_k\left(
  P_k^2 \right)$ diverges as $(\epsilon^\beta_k)^{-2}$. This
divergence corresponds to the well known phenomenon of long range
correlations in the order parameter fluctuations.

Similarly to what we did after Proposition \ref{pr:norm-mode-A}, the
proper generalisation of (\ref{eq:approx-delta}), is to consider the
case that for $k\not=0$, the support $\Delta_k$ of the measure
$d\tilde\mu_k(\lambda)$ is bounded away from $0$ and absolutely
continuous, i.e.
\begin{assumption}\label{ass:a.c.+spectr.gap}
  By translation invariance we assume that for $k\not=0$, there exists
  $\epsilon_k^\beta>0$ such that
  $\Delta_k^+\equiv\Delta_k\cap\R^+\subseteq
  [\epsilon_k^\beta,+\infty)$ and that there exists a function
  $c_k^\beta(\lambda)$ such that
\begin{equation}\label{eq:5b}
  dc_k^\beta(\lambda)=c_k^\beta(\lambda)d\lambda.
\end{equation}
\hspace*{\fill}\qed
\end{assumption}
Equation (\ref{k-correl-f-delta}) becomes
\begin{equation*}
  \lim_{n\to\infty}\omega\left( F_{n,k}(q) \hat f(H) F_{n,k}(q) \right)
  = \int_{\epsilon_k^\beta}^\infty \frac{c_k^\beta(\lambda)\lambda}
  {2(1-e^\beta\lambda)}\bigl(\hat f(\lambda)+\hat f(-\lambda)e^{-\beta\lambda}
  \bigr).
\end{equation*}
It is clear that again the single mode $(Q_k,P_k)$ gets replaced by a
continuous family of modes
$\bigl\{\bigl(Q_k(\lambda),P_k(\lambda)\bigr)\;|\; \lambda\in
\Delta_k^+\bigr\}$, such that
\begin{align*}
  \bigl[ Q_k(\lambda),P_k(\lambda')\bigr] &= c_k^\beta(\lambda)\delta(\lambda-
  \lambda')\\
  \tilde\omega_k\bigl(Q_k(\lambda)^2\bigr)&=\lambda^2 \tilde\omega_k\bigl(
  P_k(\lambda)^2\bigr)=\frac{c_k^\beta(\lambda)\lambda}{2}\coth{\beta\lambda}
  {2}\\
  \tilde\alpha_t^k Q_k(\lambda) &= Q_k(\lambda)\cos\lambda t + \lambda 
  P_k(\lambda)\sin\lambda t\\
  \tilde\alpha_t^k P_k(\lambda) &= -\frac{Q_k(\lambda)}{\lambda}\sin\lambda t 
  +  P_k(\lambda)\cos\lambda t.
\end{align*}
and
\begin{align*}
  Q_k=\int_{\epsilon_k^\beta}^\infty Q_k(\lambda)d\lambda &&
  P_k=\int_{\epsilon_k^\beta}^\infty P_k(\lambda)d\lambda.
\end{align*}

\subsection{Goldstone mode for infinite wavelength}
\label{sec:Goldstone-mode-infin-1}

Next we look for the Goldstone mode operators in the limit of $k$
tending to zero, i.e. in the long wavelength limit. We take the
results of section \ref{knot0} and study the limit $k\to 0$. Among
other results, we show that the long wavelength Goldstone mode
survives in this limit only in the ground state. This shows also that
no long wavelength quantum Goldstone modes are present for
temperatures $T>0$. For $T>0$, the spontaneous symmetry breakdown does
not show any quantum behaviour, only classical modes are present.

For simplicity we will first consider the case of a single harmonic
mode $(Q_k,P_k)$, i.e. the case (\ref{eq:approx-delta}). However we
will prove afterwards that the results we obtain in the limit $k\to 0$
are \emph{independent} of this choice and are valid in general.

Let $\epsilon_k=\lim_{\beta\to\infty}\epsilon^\beta_k$, the ground
state spectrum.  Because of the Goldstone theorem, we have that
$\lim_{k\to 0}\epsilon_k = 0$.  Let $c_0^\beta = \lim_{k\to
  0}c^\beta_k$ and $c_k=\lim_{\beta\to\infty}c^\beta_k$.
\begin{assumption}\label{assum:c-finite}
  Assume
$\lim_{k\to 0}c_k = \lim_{\beta \to \infty}c_0^\beta = c_0 < \infty$.\qed
\end{assumption}

First let $\beta < \infty$. The variances
\begin{equation*}
\tilde\omega_k\left( Q_k^2 \right)= \frac{c^\beta_k
\epsilon^\beta_k}{2}\coth\frac{\beta\epsilon^\beta_k}{2}
=(\epsilon_k^\beta)^2\tilde\omega_k\left( P_k^2 \right)
\end{equation*}
behave as follows for $k\to 0$:
\begin{align*}
  \tilde\omega_k\left( Q_k^2 \right)\approx \frac{c^\beta_k}{\beta}
  \to \frac{c_0^\beta} {\beta}\;\;\text{(finite)} &&
  \tilde\omega_k\left( P_k^2 \right) \approx \frac{c^\beta_k}{\beta
    (\epsilon^\beta_k)^2} \to \infty.
\end{align*}
Since observable fluctuation operators are always characterized by a
finite, non-zero variance, it is clear that we have to renormalize
$P_k$ before taking a limit $k\to 0$:
\begin{equation*}
\check P_k = \epsilon_k^\beta P_k.
\end{equation*}
This however implies that the commutator
\begin{equation*}
[Q_k,\check P_k]= i c^\beta_k\epsilon_k^\beta
\end{equation*}
vanishes in the limit $k\to 0$. In other words the quantum character
and hence also the harmonic oscillation of the Goldstone mode
disappears in the appropriate limit $k\to 0$, at least at non-zero
temperature.

At zero temperature ($\beta=\infty$), in the ground state, the
situation is completely different. The variances behave now for $k\to
0$ as follows:
\begin{align*}
  \tilde\omega_k\left( Q_k^2 \right) = \frac{c_k\epsilon_k}{2}\to 0 &&
  \tilde\omega_k\left( P_k^2 \right) = \frac{c_k}{2\epsilon_k}\to
  \infty,
\end{align*}
but their product
\begin{equation*}
\tilde\omega_k\left( Q_k^2 \right)\tilde\omega_k\left( P_k^2 \right)=\frac{c_k^2}
{4}\to\frac{c_0^2}{4}
\end{equation*}
remains finite. This means that the divergence of the order parameter
operator fluctuations due to long range correlations is exactly
compensated by a proportional squeezing of the symmetry generator
fluctuations.  Therefore one can find a renormalized $Q_k$ and $P_k$,
denoted by $\check Q_k$ and $\check P_k$, having both a finite,
non-zero variance, with a finite non zero commutator; indeed take e.g.
\begin{align*}
  \check Q_k \equiv \epsilon_k^{-1/2} Q_k && \check P_k \equiv
  \epsilon_k^{1/2} P_k,
\end{align*} 
then
\begin{align*}
  \tilde\omega_k\left( \check Q_k^2 \right) = 
  \tilde\omega_k\left( \check P_k^2 \right) =
  \frac{c_k}{2}\to \frac{c}{2} && \lbrack\check Q_k,
  \check P_k\rbrack = ic_k \to i c.
\end{align*}
Remark that this scaling transformation has no effect on the creation
and annihilation operators, in particular:
\begin{equation*}
a^\pm_k = \frac{Q_k \mp i\epsilon_kP_k}{\sqrt{2\epsilon_k}}
= \frac{\check Q_k \mp i\check P_k}{\sqrt{2}}.
\end{equation*}
On the other hand, the  equations of motion 
~(\ref{eq:3}) and (\ref{eq:4})
are transformed into
\begin{align*}
  \tilde\alpha_t^k \check Q_k &= \check Q_k \cos\epsilon_k t +
  \check P_k \sin\epsilon_k t\\
  \tilde\alpha_t^k \check P_k &= -\check Q_k \sin\epsilon_k t +
  \check P_k \cos\epsilon_k t.
\end{align*}
Hence in order to retain a non-trivial time evolution in the $k\to 0$ limit,
one has to rescale time as well in the following way:
$t \rightarrow \tau = \epsilon_k t$.

Let $\tilde\mathcal{B}_0$ be an algebra generated by a canonical pair
$(\check Q_0, \check P_0)$,
\begin{equation*}
  \bigl[ \check Q_0, \check P_0 \bigr] =ic_0;
\end{equation*}
$\tilde\alpha_\tau^0$, $\tau\in\R$ is a time evolution on
$\tilde\mathcal{B}_0$ defined through the equations of motion
\begin{align*}
  \tilde\alpha_\tau^0 \check Q_0 &= \check Q_0 \cos\tau + \check P_0 \sin\tau\\
  \tilde\alpha_\tau^0 \check P_0 &= -\check Q_0 \sin\tau + \check P_0 \cos\tau,
\end{align*}
and $\tilde\omega_0$ is a state on $\tilde\mathcal{B}_0$
defined through the relation
\begin{equation*}
  \tilde\omega_0\bigl(F(\check Q_0, \check P_0)\bigr)\equiv\lim_{k\to 0}
  \tilde\omega_k\bigl(F(\check Q_k, \check P_k)\bigr).
\end{equation*}
where $F$ is any polynomial in two variables.
Summarizing our results:
\begin{theorem}\label{thm:Goldstone-mode-infin}
  In the ground state ($\beta=\infty$), the dynamical system $(\tilde
  \mathcal{B}_k, \tilde\alpha_t^k,\tilde\omega_k)$ converges in the
  limit $k\to 0$ to the dynamical system
  $(\tilde\mathcal{B}_0,\tilde\alpha_\tau^0,\tilde\omega_0)$ in the
  sense that for any two polynomials $F_1,F_2$ in two variables,
  \begin{equation*}
    \tilde\omega_0\bigl(F_1(\check Q_0, \check P_0)\tilde\alpha_\tau^0
    F_2(\check Q_0, \check P_0)\bigr) = \lim_{k\to 0}
    \tilde\omega_k\bigl(F_1(\check Q_k, \check P_k)\tilde\alpha_{\frac{\tau}
      {\epsilon_k}}^k  F_2(\check Q_k, \check P_k)\bigr).
  \end{equation*}
  Therefore we can identify
  $\check Q_0 = \lim_{k\to 0} \check Q_k$ and
  $\check P_0 = \lim_{k\to 0} \check P_k$.
  Moreover $\tilde\omega_0$ is a ground state for $\tilde\alpha_\tau^0$, i.e.
  for all $X\in\tilde\mathcal{B}_0$
  \begin{equation*}
    \Bigl.\frac{d}{dt}\Bigr|_{t=0}\tilde\omega_0\bigl(X^\ast\tilde
    \alpha_\tau^0 X\bigr)\geq 0.
  \end{equation*}
  The pair $(\check Q_0,\check P_0)$ is called {\em the canonical pair
    of the collective Goldstone mode}.
\end{theorem}
\begin{proof}
  Due to quasi-freeness, it is sufficient to check these properties
  for the two-point correlation function. But in this case they follow
  immediately from the very definition of $\tilde\alpha_\tau^0$ and
  $\tilde\omega_0$. 
\end{proof}

Remark that although formally, Theorem \ref{thm:gold-boson} and
\ref{thm:Goldstone-mode-infin} are very similar,
it is  important to remember the rescaling that has been done.
In fact the previous theorem tells us that in the ground state the
long range correlations in the order parameter fluctuations are
exactly compensated by a squeezing of the generator fluctuations. Both
operators continue to form a harmonic oscillator pair in the limit
$k\to 0$, although the frequency becomes infinitesimally small and
hence the period of oscillation infinitely (or macroscopically) large.

Considering the most common case of powerlaw behaviour of the energy
spectrum, i.e. $\epsilon_k = \epsilon |k|^\delta$,
this rescaling provides information about the size of the $0$-mode
fluctuations. In a finite box $\Lambda_n$ of length $L=2n+1$, the
smallest non-zero wave vector has length $|k| \propto L^{-1}$.
Therefore the rescaling of $Q_k$ with a factor $\epsilon_k^{-1/2}$
suggests a rescaling by $L^{\delta/2}=|\Lambda_n|^{\delta/2\nu}$ of
the fluctuation, i.e.
\begin{equation*}
F_{n,0}(q)=\frac{1}{|\Lambda_n|^{\frac{1}{2}-\frac{\delta}{2\nu}}}
\sum_{x\in\Lambda_n}\bigl(q_x - \omega(q)\bigr),
\end{equation*}
in order that its variance is non-zero and finite.  This means that
the fluctuations of the symmetry generator are of order
$|\Lambda_n|^{\frac{1}{2} - \frac{\delta}{2\nu}}$, i.e. {\em subnormal
  fluctuations}.  Similarly the fluctuations of the order parameter
are of order $|\Lambda_n|^{\frac{1}{2} + \frac{\delta}{2\nu}}$, i.e.
{\em abnormal fluctuations}. This requires $\frac{\delta}{2\nu} \leq
\frac{1}{2}$, or $\delta \leq \nu$.  This condition is undoubtly
related to the condition $c < \infty$ (Assumption
\ref{assum:c-finite}). Remark also that if SSB disappears, i.e. if
$c=0$, then the Goldstone boson disappears.

Finally we remark that the results of Theorem
\ref{thm:Goldstone-mode-infin} do not depend on the particular form of
the measure $dc_k^\beta(\lambda)$, in this case given by
(\ref{eq:approx-delta}). One could equally well take the more general
form ~(\ref{eq:5b}), since in the limit $k\to 0$ this measure also
reduces to a $\delta$-peak by Proposition \ref{pr:0-delta}. It is a
straightforward calculation to show that Theorem
\ref{thm:Goldstone-mode-infin} holds in general 
(i.e. under Assumption \ref{ass:a.c.+spectr.gap}), upon interpreting
$\epsilon_k$ as the gap in the support of the measure $dc_k(\lambda)$.

Therefore we find that at zero temperature, the fluctuations of the
symmetry generator lead to a single harmonic mode with vanishingly
small frequency in the long-wavelength limit, even though at finite
wavelength, there exists a continuous family of modes associated to
the fluctuations of the symmetry generator. It is hence also
appropriate to consider the results of Theorem \ref{thm:gold-boson} as
being physically valid in general, as long as one considers low enough
temperatures and large enough wavelengths.


\begin{thebibliography}{10}

\bibitem{goldstone:1961}
Goldstone J.
\newblock {\em Il Nuovo Cimento}, 19:154, 1961.

\bibitem{kastler:1966}
Kastler D., Robinson D.W., and Swieca A.
\newblock {\em Communications in Mathematical Physics}, 2:108 -- 120, 1966.

\bibitem{swieca:1967}
Swieca J.A.
\newblock {\em Communications in Mathematical Physics}, 4:1 -- 7, 1967.

\bibitem{martin:1982}
Martin P.A.
\newblock {\em Il Nuovo Cimento}, 68 B(2):302 -- 313, 1982.

\bibitem{fannes:1982b}
Fannes M., Pul{\`e} J.V., and Verbeure A.
\newblock {\em Letters in Mathematical Physics}, 6:385 -- 389, 1982.

\bibitem{goderis:1991}
Goderis D., Verbeure A., and Vets P.
\newblock {\em Il Nuovo Cimento}, 106 B(4):375 -- 383, 1991.

\bibitem{broidioi:1991}
Broidioi M., Nachtergaele B., and Verbeure A.
\newblock {\em Journal of Mathematical Physics}, 32(10):2929 --2935, 1991.

\bibitem{broidioi:1991b}
Broidioi M. and Verbeure A.
\newblock {\em Helvetica Physica Acta}, 64:1093 -- 1112, 1991.

\bibitem{verbeure:1992}
Verbeure A. and Zagrebnov V.A.
\newblock {\em Journal of Statistical Physics}, 69:329, 1992.

\bibitem{broidioi:1993}
Broidioi M. and Verbeure A.
\newblock {\em Helvetica Physica Acta}, 66:155 -- 180, 1993.

\bibitem{goderis:1989b}
Goderis D. and Vets P.
\newblock {\em Communications in Mathematical Physics}, 122:249, 1989.

\bibitem{goderis:1990}
Goderis D., Verbeure A., and Vets P.
\newblock {\em Communications in Mathematical Physics}, 128:533 -- 549, 1990.

\bibitem{anderson:1958}
Anderson P.W.
\newblock {\em Physical Review}, 112(6):1900 -- 1916, 1958.

\bibitem{stern:1966}
Stern H.
\newblock {\em Physical Review}, 147(1):94 -- 101, 1966.

\bibitem{michoel:1999b}
Michoel T. and Verbeure A.
\newblock {\em Journal of Statistical Physics}, 96(5/6):1125 -- 1162, 1999.

\bibitem{bratteli:1996}
Bratteli O. and Robinson D.W.
\newblock {\em Operator Algebras and Quantum Statistical Mechanics 2}.
\newblock Springer Berlin, Heidelberg, New York, 1996.

\bibitem{goderis:1989}
Goderis D., Verbeure A., and Vets P.
\newblock {\em Probability Theory and Related Fields}, 82:527 -- 544, 1989.

\bibitem{arveson:1974}
Arveson W.
\newblock {\em Journal of Functional Analysis}, 15(3):217 -- 243, 1974.

\bibitem{landau:1981}
Landau L., {Fernando Perez J.}, and Wreszinski W.F.
\newblock {\em Journal of Statistical Physics}, 26(4):755 -- 766, 1981.

\bibitem{wreszinski:1987}
Wreszinski W.F.
\newblock {\em Fortschritte der Physik}, 35(5):379 -- 413, 1987.

\bibitem{thirring:1967}
Thirring W. and Wehrl A.
\newblock {\em Communications in Mathematical Physics}, 4:303 -- 314, 1967.

\bibitem{thirring:1968}
Thirring W.
\newblock {\em Communications in Mathematical Physics}, 7:181 -- 189, 1968.

\bibitem{requardt:1982}
Requardt M.
\newblock {\em Journal of Statistical Physics}, 29(3):117 -- 127, 1982.

\bibitem{bogoliubov:1962}
Bogoliubov N.N.
\newblock {\em Phys. Abh. S.U.}, 1:229, 1962.

\bibitem{michoel:1998}
Michoel T., Momont B., and Verbeure A.
\newblock {\em Reports on Mathematical Physics}, 41(3):361 -- 395, 1998.

\bibitem{narnhofer:1983}
Narnhofer H., Requardt M., and Thirring W.
\newblock {\em Communications in Mathematical Physics}, 92:247 -- 268, 1983.

\end{thebibliography}
\end{document}